\documentclass[%
 reprint,
 amsmath,amssymb,
 aps,
]{revtex4-2}

\usepackage{graphicx}
\usepackage{dcolumn}
\usepackage{bm}

\begin{document}


\title{Studies of Beam Intensity Effects in Fermilab Booster Synchrotron}

\author{Jeff Eldred, Valeri Lebedev, Kiyomi Seiya, Vladimir Shiltsev}
 \email{shiltsev@fnal.gov}%
 \affiliation{Fermi National Accelerator Laboratory, PO Box 500, MS339, Batavia, IL 60510,USA}




\date{\today}
\begin{abstract}
Detrimental beam dynamics effects limit performance of high intensity rapid cycling synchrotrons (RCS) such as the 8 GeV proton Fermilab Booster. Here we report the results of comprehensive experimental studies of various beam intensity dependent effects in the Booster. In the first part, we report the dependencies of the Booster beam intensity losses on the total number of protons per pulse and on key operational parameters such as the machine tunes and chromaticities. Then we cross-check two methods of the beam emittance measurements (the multi-wires proportional chambers and the ionization profile monitors). Finally we used the intensity dependent emittance growth effects to analyze the ultimate performance of the machine in present configuration, with the maximum space-charge tuneshift parameter $\Delta Q_{SC}\sim 0.6$, and after its injection energy is upgraded from 0.4 GeV to 0.8 GeV.
\end{abstract}

\maketitle

\section{INTRODUCTION}
\label{Intro}

High-energy high-intensity proton beams are widely used for a broad spectrum of scientific research, including particle colliders, muon spectroscopy, crystallography of biological molecules, neutrino physics, and neutron scattering. Over the past four decades, the power of such beams has increased by about three orders of magnitude with average doubling time of about four  years \cite{shiltsev2011complex, shiltsev2020PhysToday}. Growing demands of the physics research call for more powerful, more productive and more sophisticated proton accelerators. For example, at present, the leading accelerators for neutrino research are rapid cycling synchrotron (RCS) facilities J-PARC in Japan which has reached 515 kW of the 30 GeV proton beam power, and the Fermilab Main Injector delivering up to 766 kW of 120 GeV protons, while the physics needs call for the next generation, higher-power, megawatt and multi-MW-class {\it superbeams facilities} \cite{shiltsev2020superbeams}. 

Further progress in the proton beam power needs improvements in the accelerator technology \cite{wei2003synchrotrons} and addressing problems related to desired beam pulse structure, beam losses, and the lifetimes of beam targets \cite{zwaska2018targets, simos2019target, sola2019beam}. Of particular challenge are issues associated with the beam dynamics, such as space-charge effects, instabilities, emittance growth, halo formation and losses, collimation, accumulation of secondary charges, linear and non-linear optics optimizations, etc - see comprehensive analysis in Refs.\cite{chao1993physics, diskansky1997physics,  ng2006physics, Reiser, hofmann2017space}. Advanced beam instrumentation, analytical methods and diagnostics \cite{minty2003measurement, strehl2006beam, huang2006emittance, wittenburg2013instrum}, and experimental beam studies are critical for adequate understanding of the beam dynamics.     

Here we present results of the experimental studies of high intensity beam dynamics in the Fermilab 8 GeV proton Booster RCS, carried out as part of the Summer 2019 Booster Studies program \cite{eldred2019physics}. In Sec.\ref{IntroBooster} we give a brief description of the accelerator. Next Sec.\ref{BeamLosses} is devoted to the beam loss diagnostics and measurements of the machine transmission efficiency and its dependence on the total number of protons per pulse, tunes and chromaticities. The beam emittance diagnostics and measurements, including analysis of the intensity dependent emittance growth, are presented in Sec.\ref{Emittance}. Finally, we discuss general scaling laws and ultimate performance of the machine now and after upcoming upgrade in Sec.\ref{Discussion}.  

\begin{figure}[htbp]
\centering
\includegraphics[width=0.99\linewidth]{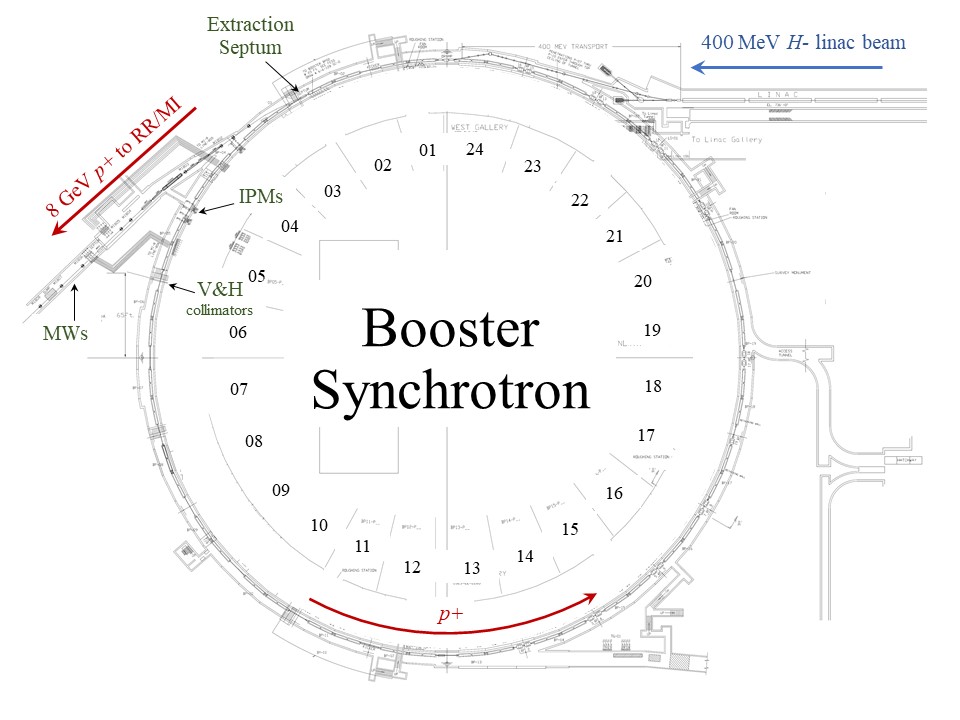}
\caption{Schematics of the Fermilab Booster synchrotron. Sectors, each consisting of four combined function magnets, are numerated 1 to 24. Indicated are locations of the Ionization Profile Monitors (IPMs), Multi-Wire (MWs) beam profile monitors in the 8 GeV proton transport line 
and the 400 MeV $H^-$ beam injection line.} 
\label{FigBooster}
\end{figure}

\section{FNAL BOOSTER SYNCHROTRON}
\label{IntroBooster}

The complex of Fermilab proton accelerators includes a 750 keV $H^-$ RFQ, 400 MeV $H^-$ pulsed normal-conducting RF linac, 8 GeV proton Booster synchrotron, 8 GeV Recycler storage ring that shares tunnel with 120 GeV proton Main Injector synchrotron, and a 3.1 GeV muon Delivery Ring \cite{ShiltsevMPLA, Convery}. About 16 km of beamlines connect the accelerators, bring the beams to fixed targets and to experiments for high energy particle physics research at the Intensity Frontier. There are plans to further increase the facility power from the current world leading level of $\sim$750 kW of average 120 GeV beam power on the neutrino target to over 1.2 MW at the start of the LBNF/DUNE experiment \cite{dune2016lbnf} in the second half of the 2020’s via replacement of the existing 
400 MeV normal-conducting Linac with a modern CW-capable 800 MeV superconducting RF linear accelerator (PIP-II, see \cite{PIPII2017lebedev}) and corresponding beamline for injection into the Booster. There are also several concepts to further double the beam power to $>$2.4 MW after replacement of the existing 8 GeV Booster synchrotron \cite{eldred2019pipiii, eldred2020snowmass}.
\begin{table}[h]
\begin{center}
{\begin{tabular}{lccc}
\hline
Parameter & & Comments \\
\hline 
Circumference, $C$ & 474.20 m & \\
Inj.energy (kin.), $E_i$	& 400 MeV	& $\beta_i$=0.701, $\gamma_i$=1.426 \\
Extr.energy (kin.), $E_f$	& 8 GeV	& $\beta_f$=0.994, $\gamma_f$=9.526 \\
Cycle time, $T_0=1/f_0$ & 	1/15 s & 20,000 turns \\
Harmonic number, $h$ &	84	& \\
RF frequency, $f_{RF}$& 37.77-52.81 MHz & inj.-extr.	\\
Max. RF voltage, $V_{RF}$ &1.1 MV	& \\
Trans. energy, $E_{tr}$ &	4.2	GeV	& $\gamma_{tr}$=5.478, at $t$=17ms \\
No. of cells, magnets	& 24, 96 & $FOFDOOD$, 96$^o$/cell \\
Total intensity, $N_p$	& 4.5$\cdot 10^{12}$ & $N_b$=81 bunches \\
Rms norm.emitt., $\varepsilon_{x,y}$ & 2.0 $\pi \,\mu$m	& 12 $\pi \, \mu$m for 95 \% \\
$\beta$-functions, $\beta_{x,y}$ &	33.7/20.5 m & 	max.  \\
Dispersion, $D_x$ &	3.2	m & 	max.  \\
\hline
\end{tabular}}
\caption{Main operational parameters of the Fermilab Booster.}
\label{BoosterParam}
\end{center}
\end{table}

\begin{figure}[htbp]
\centering
\includegraphics[width=0.99\linewidth]{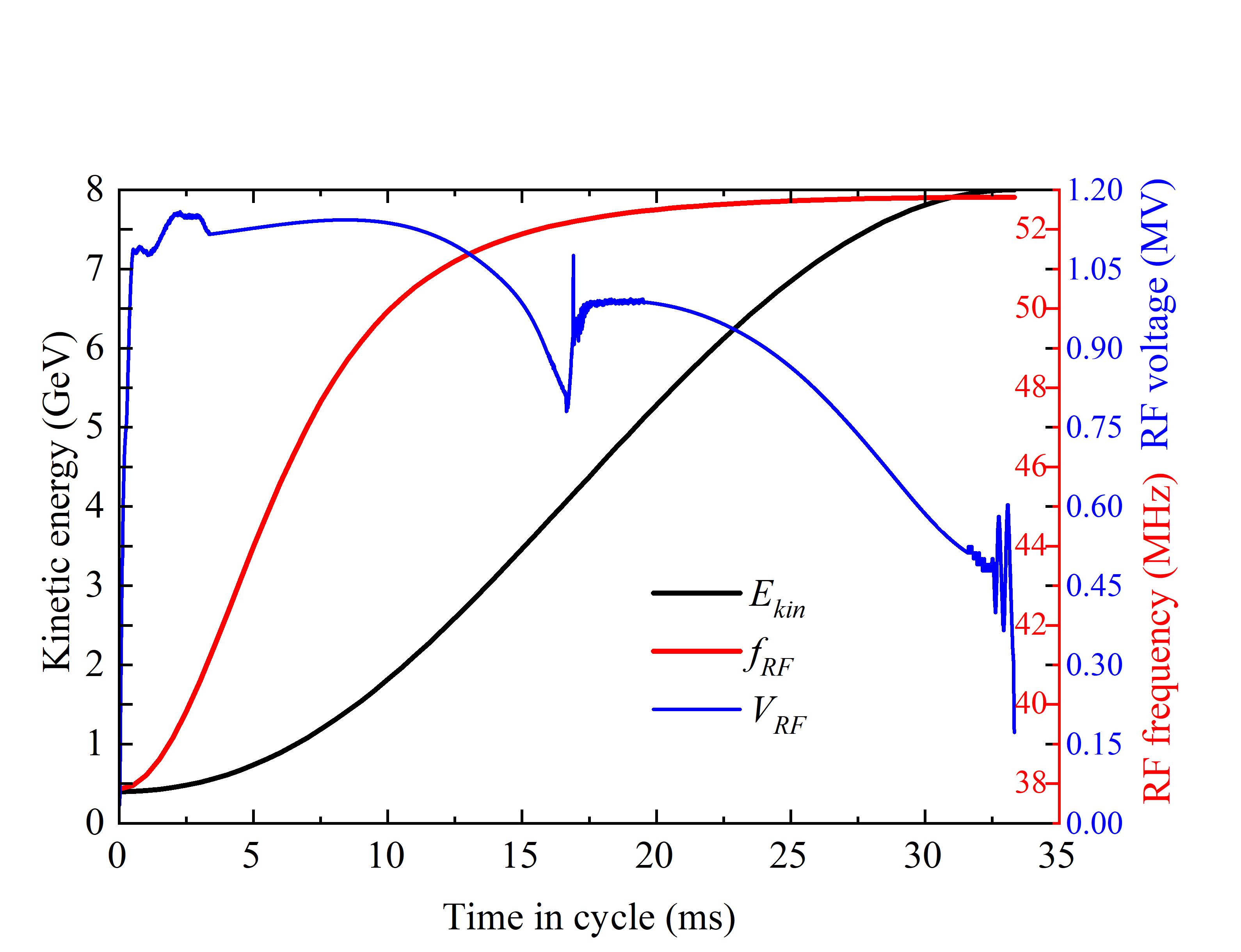}
\caption{Booster ramp: kinetic energy (black), RF voltage(blue) and frequency (red).} 
\label{FigBoosterRamp}
\end{figure}

The Fermilab Booster \cite{Booster} is a 474.2 m circumference, alternating-gradient, rapid-cycling synchrotron containing 96 combined-function magnets – see Fig.\ref{FigBooster}. Together with capacitor banks, these magnets form a resonant network ($Q=40$) and get excited with a 15-Hz biased sine wave. Beam acceleration ramp from 0.4 GeV at injection to 8.0 GeV at extraction is 33.3 ms long - half of the magnet cycle period – and contains about 20 000 turns. Correspondingly, all the parameters of the machine and beam significantly vary in the cycle – from the currents in all correctors (trim dipoles, trim quads and skew quads, sextupoles and octupoles) to RF frequency, voltage and phase (see Fig.\ref{FigBoosterRamp}), from the betatron and synchrotron tunes and chromaticities to proton beam intensity, positions, sizes, emittances, bunch length and energy spread, etc. Main parameters of the Booster are given in Table \ref{BoosterParam}. Without going into details of the Booster high intensity operation and interface with other machines in the complex which can be found in Refs.\cite{BoosterBook, eldred2015phd, seiya2015beam}, here we only briefly outline main processes which occur at injection, transition crossing and extraction. 

\begin{figure}[htbp]
\centering
\includegraphics[width=0.99\linewidth]{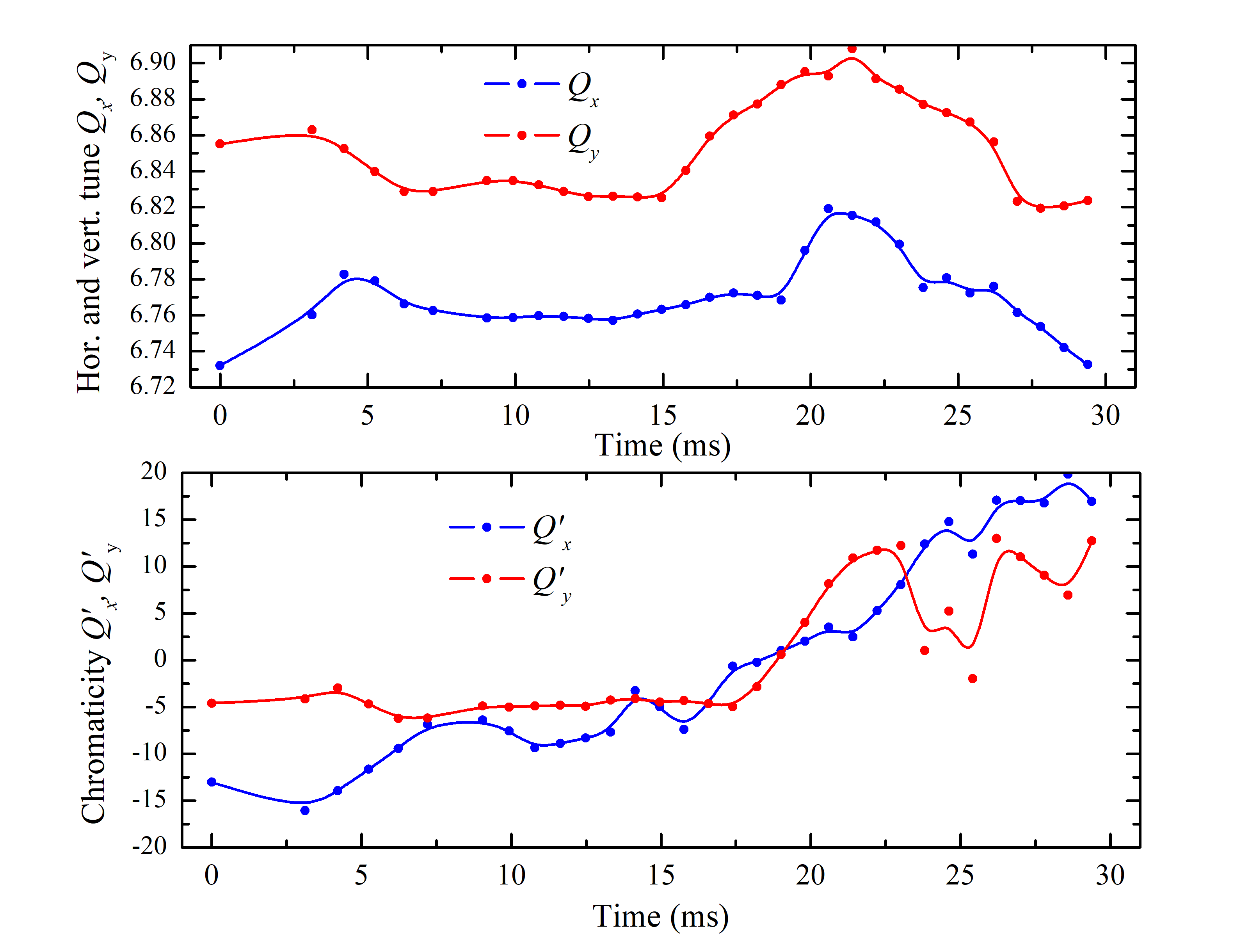}
\caption{Measured Booster beam characteristics during the acceleration cycle: (top) the horizontal (blue) and vertical (red) betatron tunes $Q_{x,y}$; (bottom) (top) the horizontal (blue) and vertical (red) chromaticities $Q'_{x,y}$.} 
\label{FigBoosterTunes}
\end{figure}

The Booster receives 400 MeV $H^-$ beam of 201 MHz bunches from the Linac while it is close to minimum of the magnetic field ramp. $H^-$ particles are stripped of two electrons when pass through a thin foil and resulting protons are accumulated over many turns in the ring (the scheme known as charge exchange injection \cite{budkerdimov1963stripping, hojvat1979multiturn}). Correspondingly, the total injected and accelerated beam intensity $N_p$ scales with the Linac current, which is typically $\sim$25 mA, and the total number of injection turns $N_{turns}$, approximately as  $N_p=0.34 \cdot 10^{12} \cdot N_{turns}$, e.g., about $4.8 \cdot 10^{12}$ for a typical 14-turn injection. The duration of the beam injection also scales with the number of turns as 2.2$\cdot N_{turns}$ [$\mu$s]. The beam is injected with RF voltage close to zero and then adiabatically captured over about 300 $\mu$s by the RF system \cite{bhat2015newinj, bhat2017injstudies}. Right after injection and in the following several milliseconds, the high intensity protons beam is subject of the strongest space-charge forces, characterized by the space-charge parameter $\Delta Q_{SC} \geq 0.5$ (see below). Transverse and longitudinal ring impedances are large \cite{ng2006physics, macridin2011coupling, macridin2013transverse} and the Booster operation requires simultaneous, fast and idiosyncratic adjustment of orbits, optical functions \cite{valishev2016suppression}, tunes, chromaticities and many other machine parameters on top of changing energy, RF voltage and frequency (see Fig.\ref{FigBoosterTunes}). 

The transition occurs at about 17 ms into the Booster cycle (after injection), at the beam energy of 4.2 GeV. The Booster is currently operated without a dedicated $\gamma_t$-jump system, though the current of trim quadrupoles, RF system voltage and RF frequency curves have been tuned to minimize the losses and control the longitudinal emittance, which somewhat grows from its initial 95\% value of about 0.08 eVs \cite{yang2005trans, lebedev2016trans, ostiguy2016trans}. 


The rapid acceleration in the Booster requires large accelerating voltage. 
To inject the Booster beam efficiently into the Recycler for slip-stacking, it is desirable to rotate the beam in longitudinal phase-space so to reduce the momentum spread \cite{yang2007momentum}. The Booster bunch rotation is performed via quadrupole excitation of the synchrotron oscillation as the RF voltage is modulated at twice the synchrotron frequency and this drives a longitudinal quadrupole resonance. Once the beam energy is close to the extraction energy we perform snap bunch rotation, i.e., at about 2 ms before the end of the cycle, the RF voltage is increased slowly to 650 kV to increase the energy spread of the bunches and dropped down rapidly to 130 kV. This gives the required small energy spread for the beam for slip stacking in the downstream accelerators.

 \begin{figure}[htbp]
\centering
\includegraphics[width=0.99\linewidth]{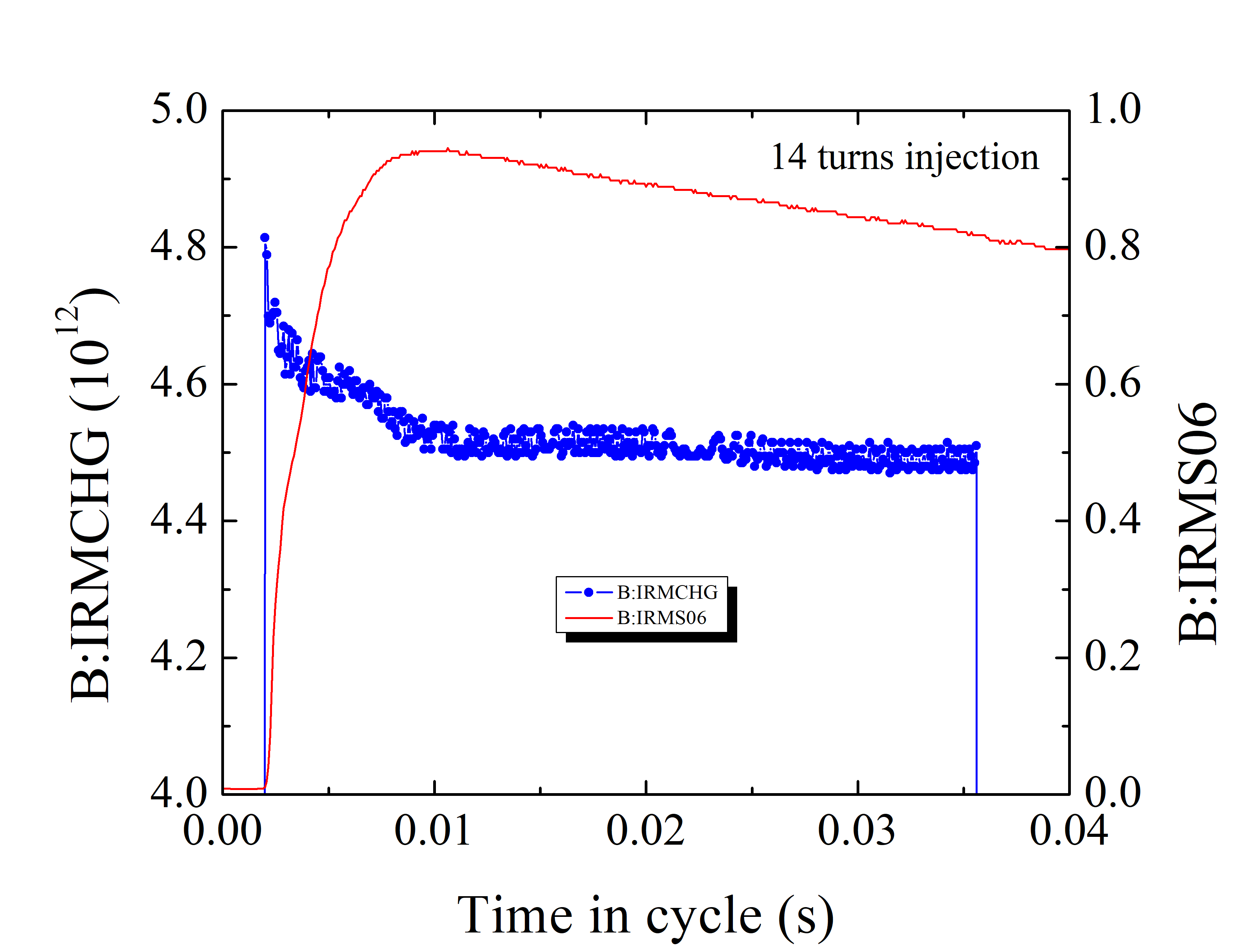}
\includegraphics[width=0.99\linewidth]{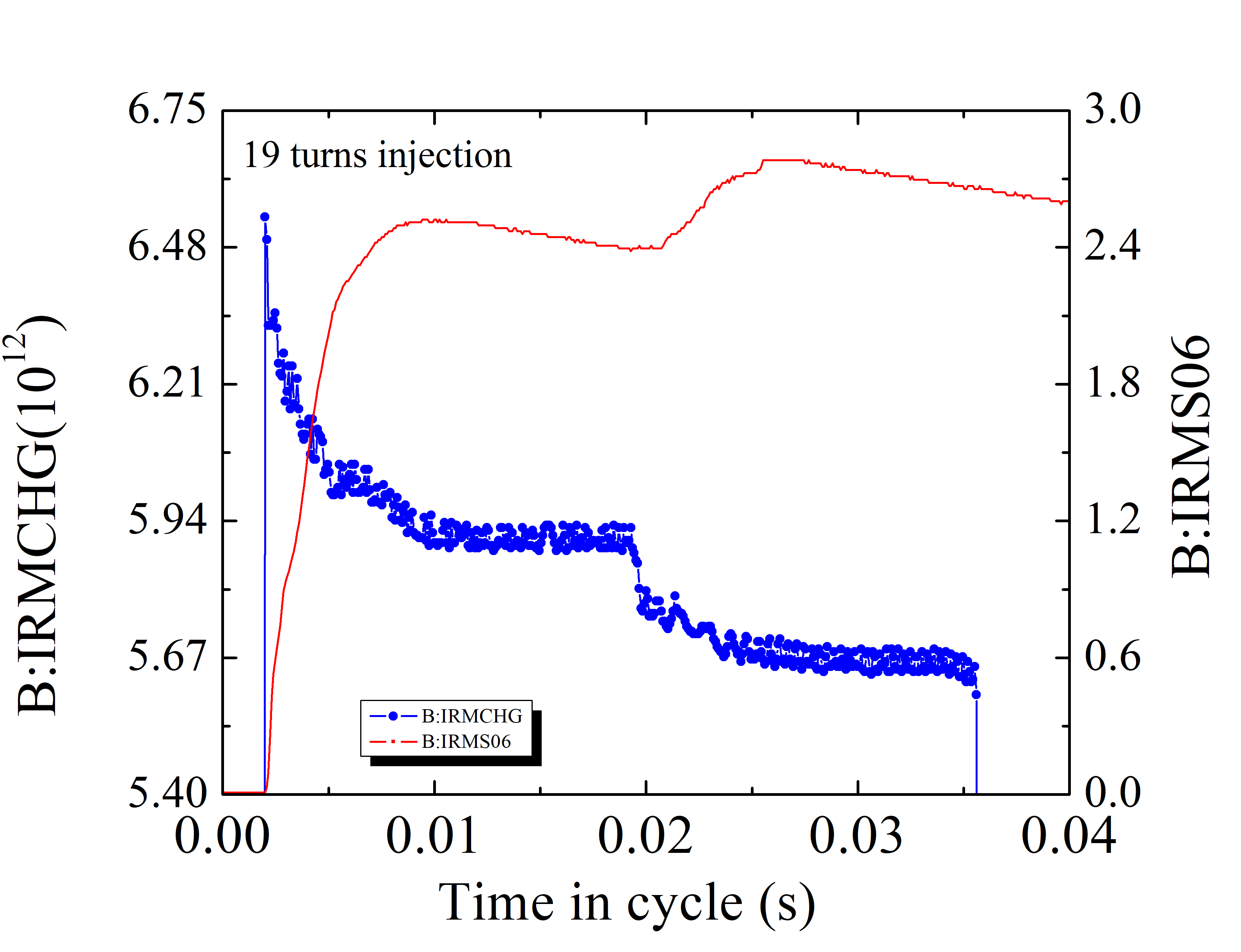}
\caption{
Booster beam losses in the acceleration cycle: (top) the BCHG0 intensity monitor signal (blue, left axis) and the S06 BLM readings (red, right axis) for nominal operational 14 turns injection intensity; (bottom) the same for higher intensity 19 turns injection. }
\label{FigBoosterBLMsignals}
\end{figure}

The overall average Booster beam loss limit has been administratively set to $W$=525W, i.e., 35J per cycle with 15 Hz beam cycles. Such level allows us to maintain all elements in the Booster tunnel without excessive radiation exposure and corresponds to either 13\% beam loss at injection energy or 1.2\% at the transition, or 0.6\% at extraction for nominal intensity of about 4.5$\cdot 10^{12}$ protons per pulse. As illustrated in Fig.\ref{FigBoosterBLMsignals}, at nominal intensity, with 14 turns injection, the losses mostly occur at injection, but the loss at transition becomes dominant at higher intensities. In general, the beam loss induced radiation is the most important and most challenging factor limiting the performance of high intensity RCSs \cite{wei2003synchrotrons}.

\section{BEAM LOSSES}
\label{BeamLosses}

\subsection{Beam loss diagnostics }
\label{LossDiagnostics}

The main diagnostic of the total circulating Booster beam intensity is the ACNET~\cite{acnet2019} (Fermilab global accelerator control system) channel B:CHG0, that uses the signal from the beam current  toroid properly mixed and averaged with the RF waveform signal. 
The raw B:CHG0 toroid intensity data are quite reliable and accurate to better than a fraction of one percent at relatively stable machine and beam parameters, but require correction of the systematic effects early in the cycle, when the RF frequency, the bunch length and the bunch structure are quickly changing in time. The required correction was established using a cross calibration of the intensity loss $\Delta N_p$ reported by B:CHG0 and the power loss signal measured by the most appropriate beam loss monitor BLMS06 - see Fig.\ref{FigBoosterBLMsignals}. The BLM at S06 is located between two Booster collimators, and in regular operation registers the most significant loss flux over the cycle. The BLM signal represents an integral of a proportional chamber with decay time of $\sim$200 ms. This signal is reset at the beginning of each cycle. As one can see BLMS06 peaks for the first time at 10 ms into the cycle with the amplitude proportional to the power loss integrated well over the initial lossy period of the Booster cycle. The loss monitor signals at 10 ms are compared to the  reported changes in the toroid signal B:CHG0 and it was found that the latter overestimates $O$(5\%) losses at small intensities by about 2\% but that difference disappears (mutual linearity gets restored) at higher intensities and higher fractional losses - see details in \cite{Shiltsev2020a}.  

\begin{figure}[htbp]
\centering
\includegraphics[width=0.99\linewidth]{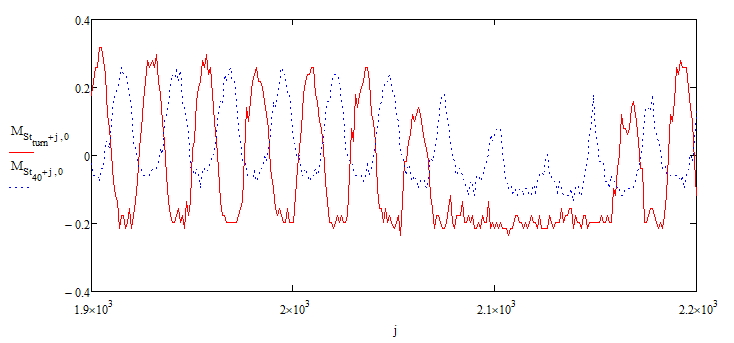}
\caption{Booster RW monitor traces for the bunch beam current profiles right before (dashed blue) and 40 turns after (solid red) the extraction gap clearing.}
\label{FigBoosterGap}
\end{figure}

\subsection{Fractional beam loss}
\label{FractionalLoss}

Our studies reveal three kinds of the Booster beam intensity losses : 1) minor intensity-independent loss at injection $\sim$(1-2)\% due to the so-called 
"notch clearing"; 2) significant $O(5\%)$ loss shortly after the injection that can be attributed to space-charge effects as it strongly depends on the intensity, the working point tunes and chromaticities; and 3) threshold-like increase of the losses $O(5\%)$ during and after the transition  above $N_p=6 \cdot 10^{12}$.


\textit{The notch clearing.} The Booster extraction kicker rise time is about 70 ns long, and to reduce the losses at extraction one requires an empty gap in the beam structure. The beam gap is created by removing 3 out of 84 possible bunches at the lowest feasible beam energies. First, a recently built laser system \cite{johnson2018mebt} is used to create the notch within a Linac beam pulse, immediately after the RFQ at 750 keV, where activation issues are absent. The beam with such a gap is injected into the Booster. The laser notch system is not 100\% efficient and, in addition, some particles slip into the gap before the injection RF capture and as the result there is 400 MeV beam in the gap – as illustrated by Fig.\ref{FigBoosterGap}. These particles are cleared out at approximately 150 turns after the injection by a kicker pulse that removes 1.4$\pm$0.4\% of the total beam intensity. The particles are directed to a special in-line beam dump which intercept them with efficiency much better than the efficiency of the collimation system, so these losses are excluded from the following analysis of the intensity dependent effects. Ongoing improvements to the laser notcher power and laser interaction cavity are expected to greatly reduce the number of unwanted particles in the notch gap.  

\begin{figure}[htbp]
\centering
\includegraphics[width=0.99\linewidth]{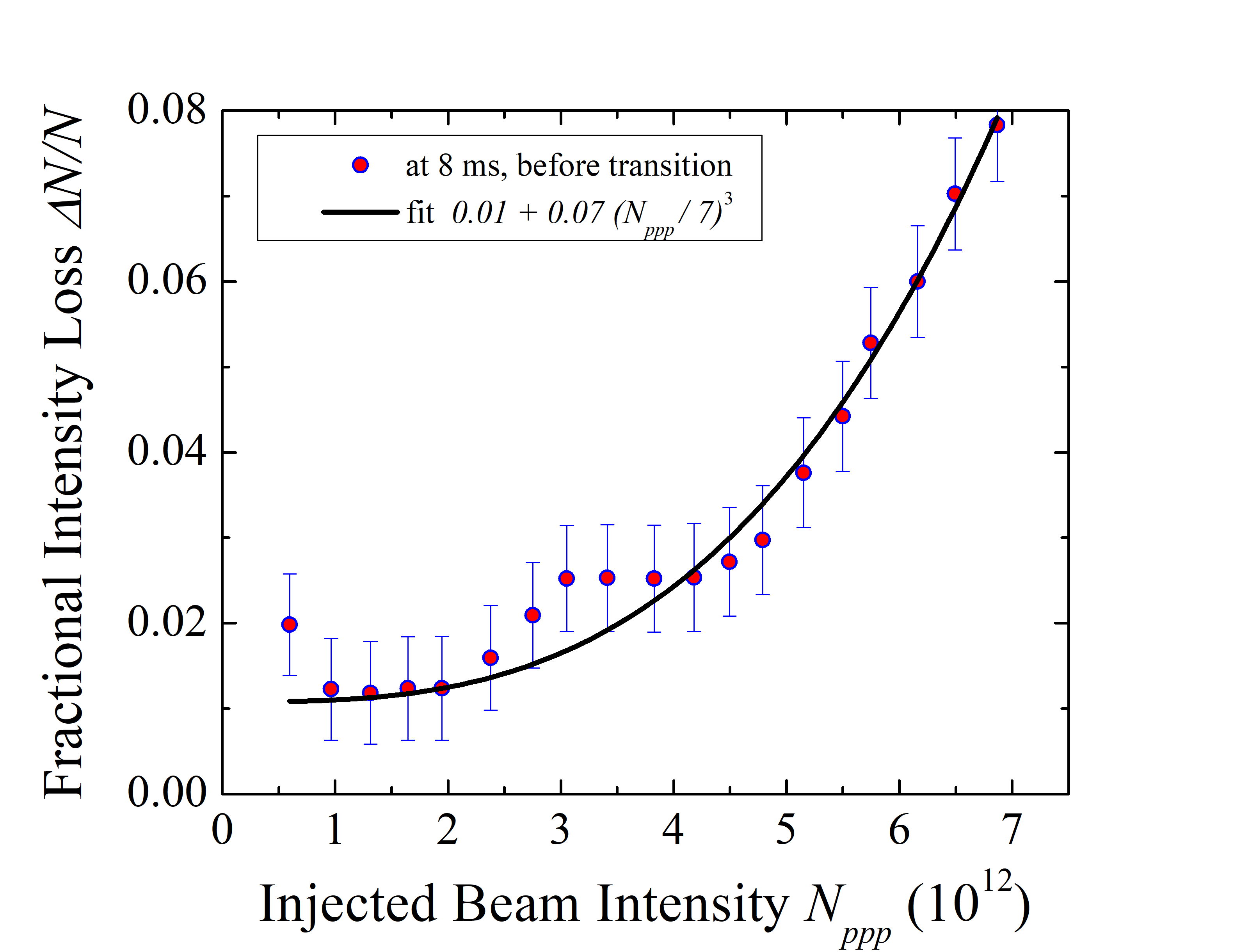}
\caption{Intensity-dependent fractional Booster beam intensity loss at injection vs total number of protons.  }
\label{FigBoosterFractLoss}
\end{figure}

The largest, and of the most operational concern,  \textit{intensity dependent losses} which take place over the first 8 ms after the injection are presented in Fig.\ref{FigBoosterFractLoss}. As one can see, the losses quickly grow with $N_p$ – solid line in Fig.\ref{FigBoosterFractLoss} is for the fit
\begin{equation}
\frac{\Delta N_p}{N_p} = 0.01 + 0.07 \cdot \Big( { N_p \over {7 \cdot 10^{12}}} \Big)^3 \, .    
\label{EQ1}
\end{equation}
The non-zero intercept at small $N_p$ might indicate either a different mechanism of constant losses at low intensity or an insufficient measurement accuracy. 

The beam intensity losses at the {\it transition crossing} were found to be below $\lesssim 0.5$\% up to a threshold at about $N_p=6 \cdot 10^{12}$ and then quickly reach 7\% at $N_p=7 \cdot 10^{12}$ \cite{Shiltsev2020a}. The characteristic dependence and underlying physics mechanism is quite different from what is observed at the injection. The intensity dependence at transition is most likely related to the longitudinal beam loading and the voltage induced by large longitudinal impedance of the Booster laminated vacuum chamber interior \cite{burov2012laminated, macridin2011coupling, lebedev2016trans}. We also have to note that for optimal transmission efficiency, the transition crossing has to be retuned for higher intensities, while during our studies we operated with the RF and other machine parameters at the transition optimized for nominal operational intensity of $N_p=4.5 \cdot 10^{12}$ . 


Altogether, in our study the fractional beam intensity losses over the entire Booster cycle are about $4.2\pm0.5$\% for the nominal intensity while at the record high injected intensity $N_p=7 \cdot 10^{12}$ they are about 15\%. Obviously, such losses are not acceptable for routine operation within the administrative beam power loss limit. 

\subsection{Tune scans}
\label{Tunescans}

\begin{figure}[htbp]
\centering
\includegraphics[width=0.99\linewidth]{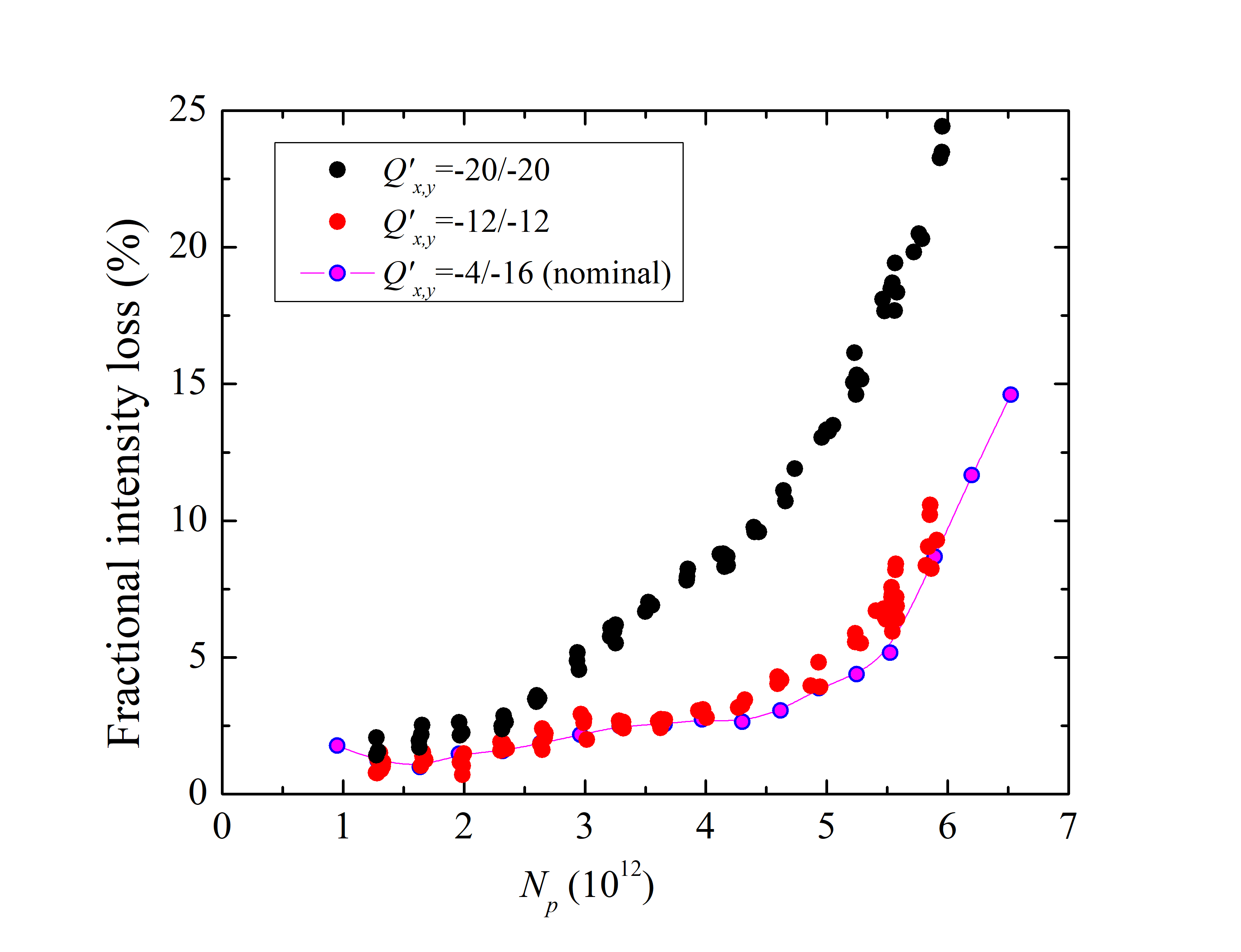}
\caption{Transmission efficiency 1 ms after injection for different chromaticities $Q'_{x,y}$ vs $N_p$.}
\label{FigChromLoss}
\end{figure}

To better understand the nature of the intensity loss phenomena we have studied the Booster transmission efficiency dependence on the chromaticities and tunes -   the $Q'_{x,y}$ and $Q_{x,y}$ scans.  Fig.\ref{FigChromLoss} shows the dependence of the losses over the first 1 ms of the Booster cycle ($\sim$450 turns) at the nominal operational tunes $Q_{x,y}$= 6.78/6.88 but at three different chromaticity settings – the nominal one $Q’_{x,y}$=-4/-16, and then at $Q’_{x,y}$=-12/-12, and at $Q’_{x,y}$=-20/-20. The fractional losses were calculated out of the B:CHG0 signal, corrected for the systematic error at lower intensities and with the extraction gap clearing loss subtracted, following the method presented in the preceding section. 

The results presented in Fig.\ref{FigChromLoss} clearly show significant increase of the losses with the chromaticity. Taking for simplicity the same functional dependence on intensity as in Eq.(\ref{EQ1}), the chromatic dependence of the loss data is consistent with
\begin{equation}
\frac{\Delta N_p}{N_p} = 0.013+0.10 \cdot \Big( { N_p \over {7\cdot10^{12}}} \Big)^3 \Big({Q'_{avg} \over 10}\Big)^{1.9\pm 0.2} ,    
\label{EQ2}
\end{equation}
with  $Q’_{avg}=(|Q'_y|+ |Q'_y|)/2$ denoting the average chromaticity. There is a lower limit on operational chromaticity that depends on the intensity and is usually associated with the need to maintain the coherent beam stability. Correspondingly, the low chromaticity operation is possible only at low intensities. Notably, the strong dependence of the losses at injection on the chromaticity, presumably due to the space-charge effects, $\Delta N_p \propto Q’^2$, is similar to the incoherent beam losses due to parasitic beam-beam effects observed in the Tevatron collider \cite{shiltsev2005beambeam}. 

\begin{figure}[htbp]
\centering
\includegraphics[width=0.49\linewidth]{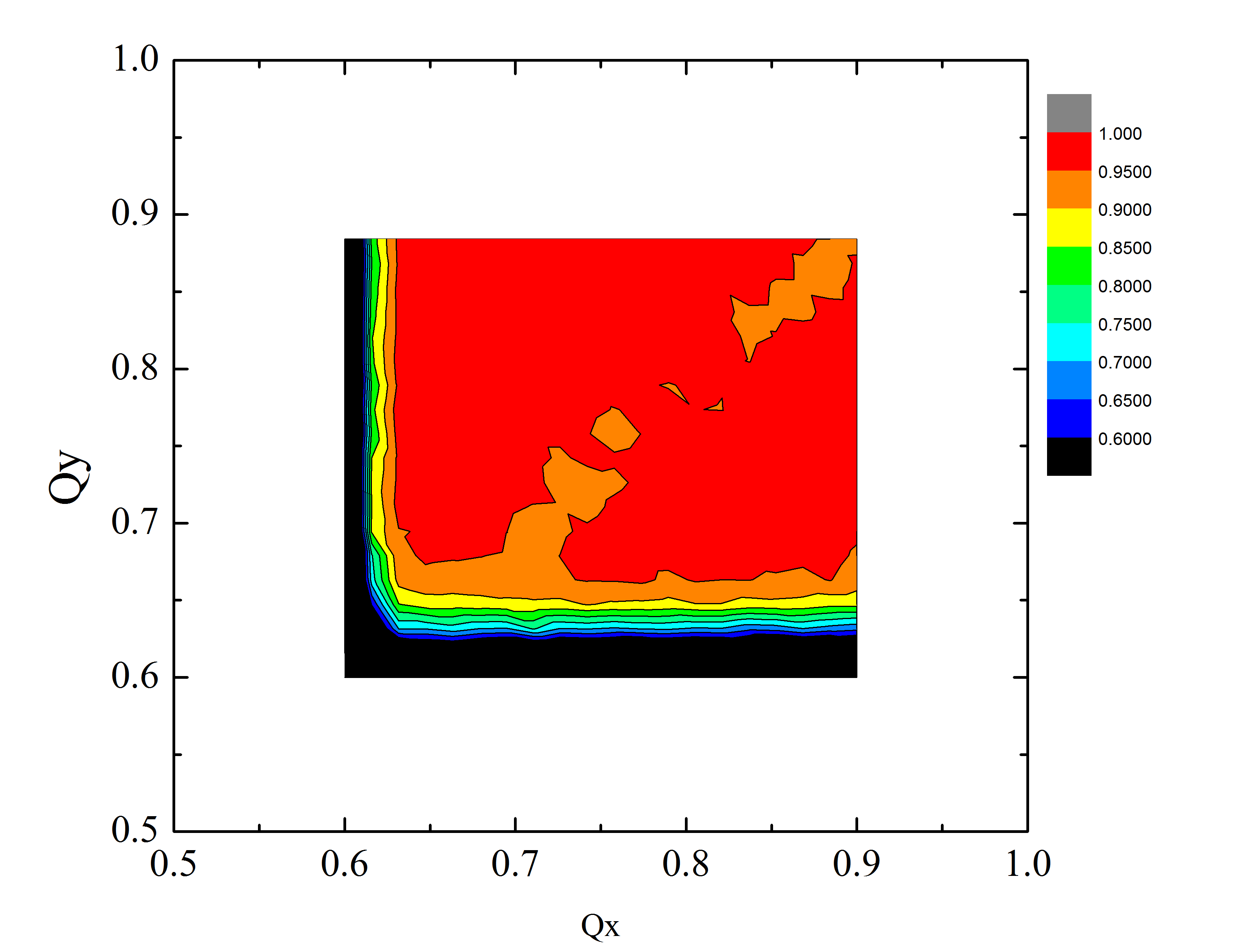}
\includegraphics[width=0.49\linewidth]{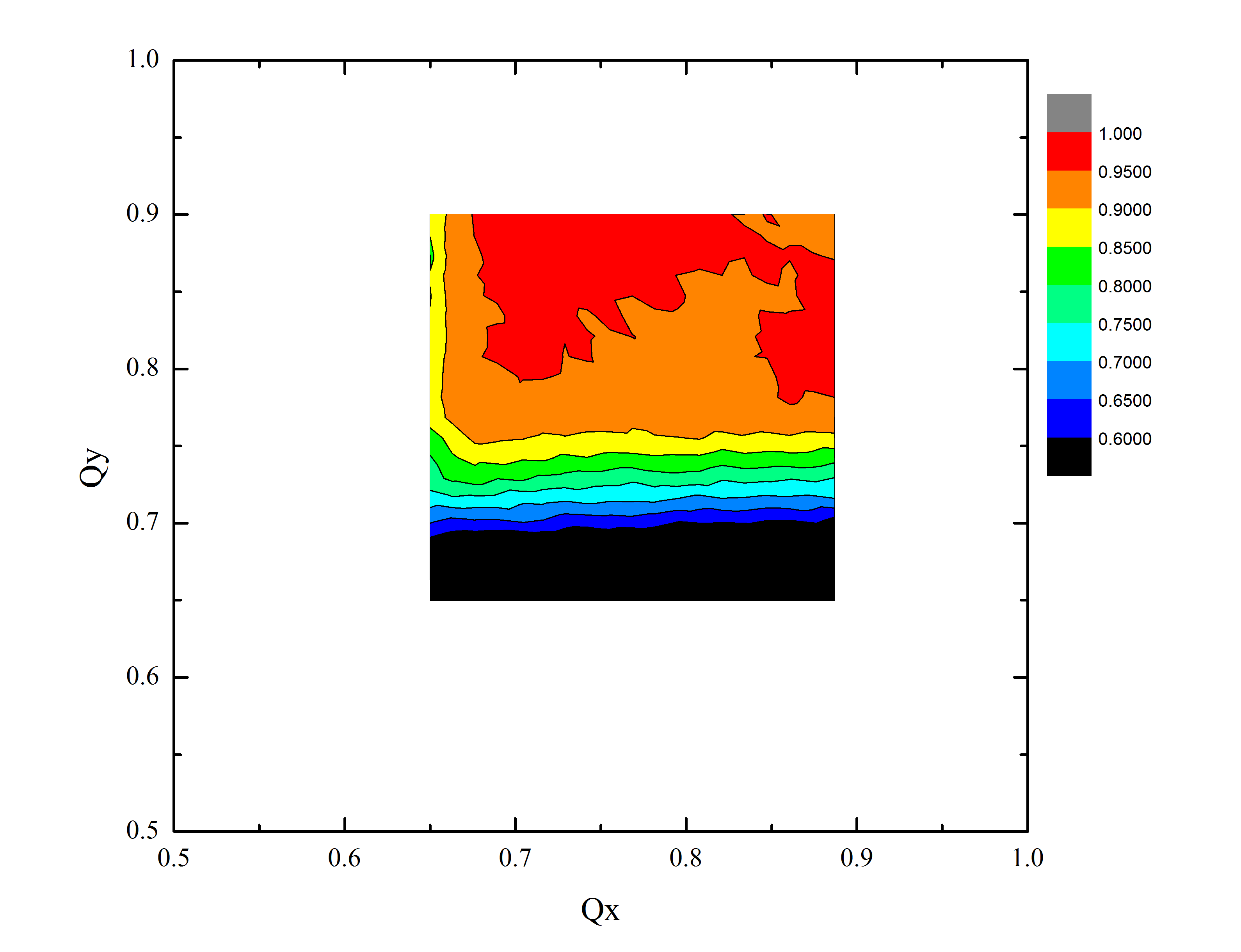}
\includegraphics[width=0.49\linewidth]{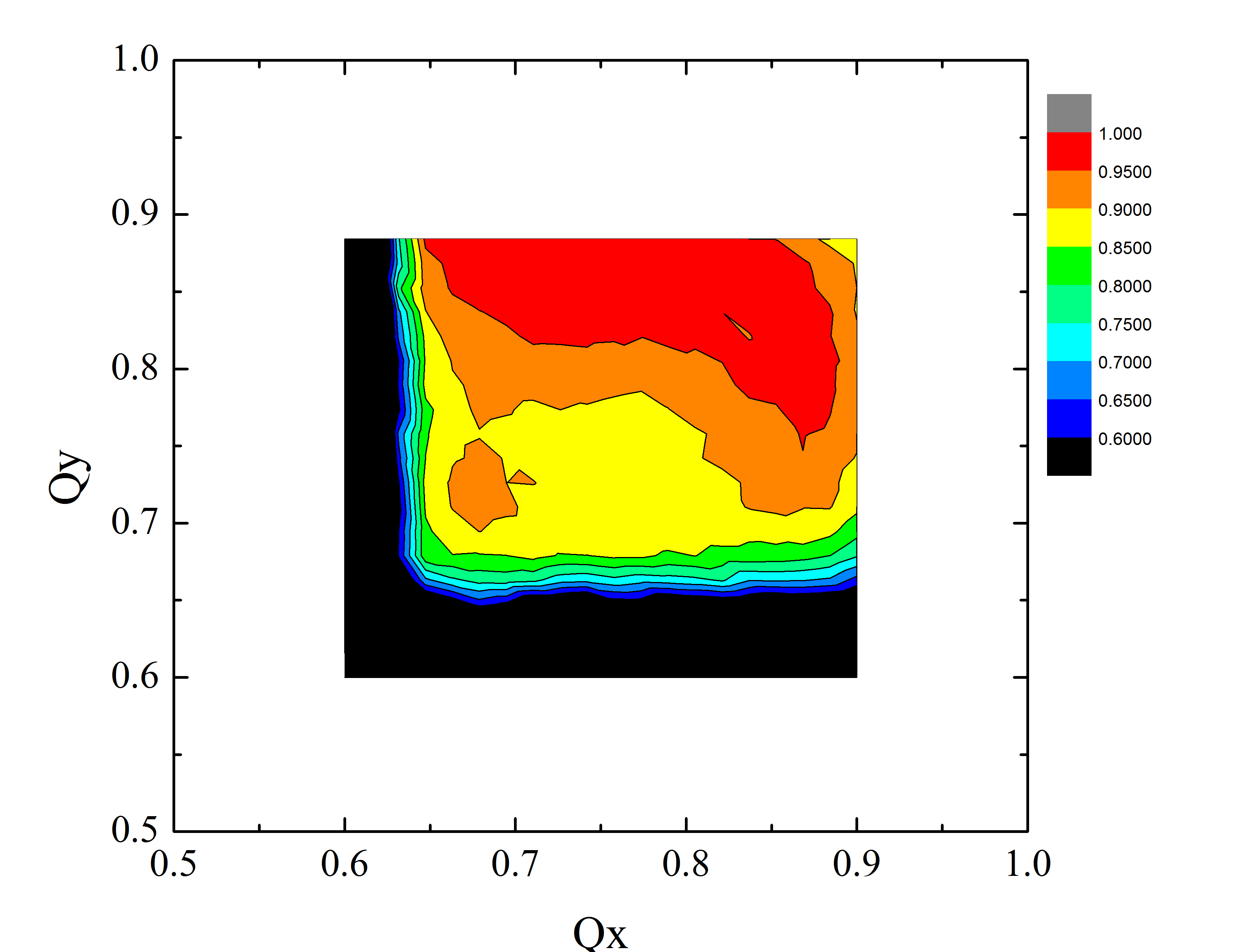}
\includegraphics[width=0.49\linewidth]{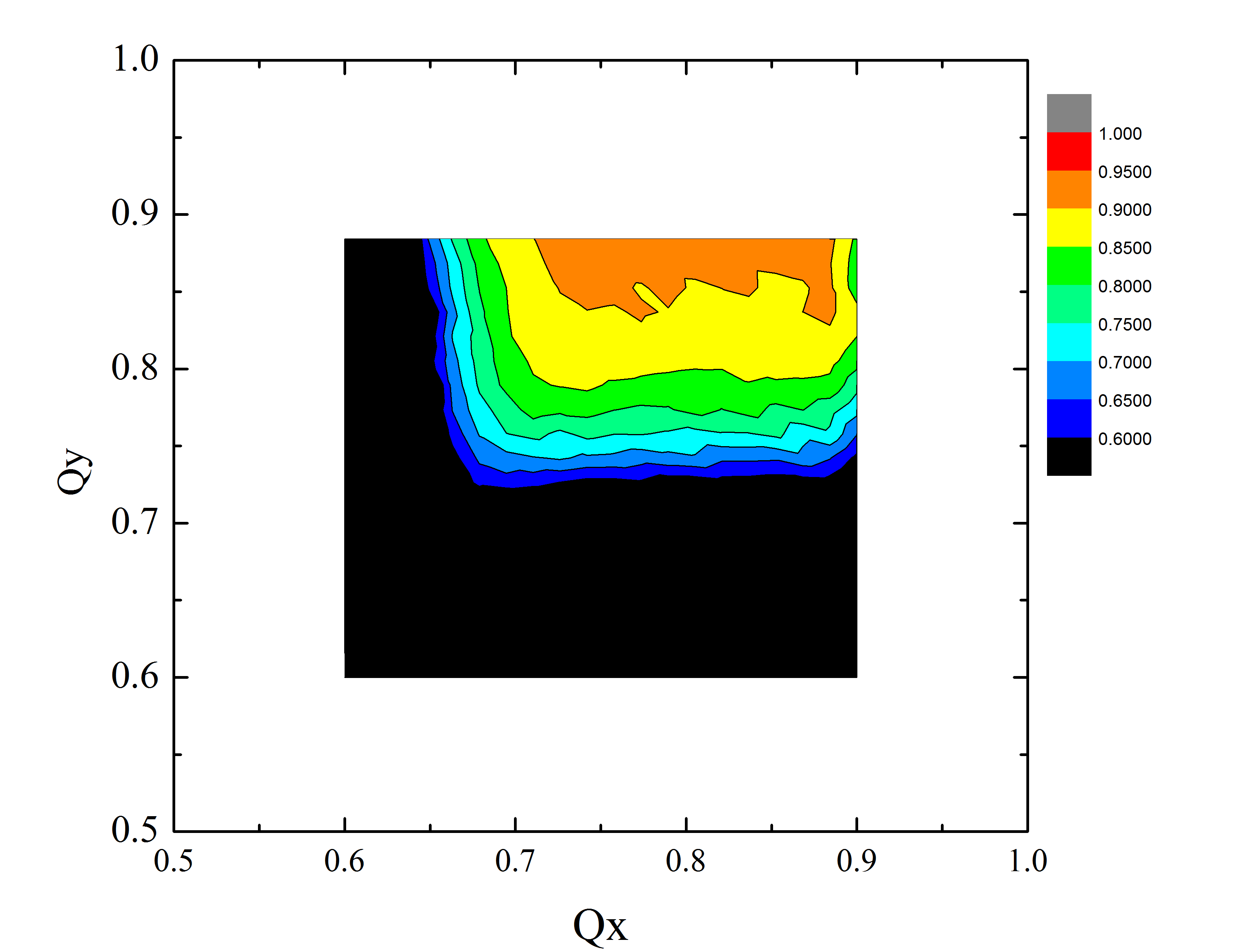}
\caption{Tune scans of the transmission efficiency over the first millisecond after injection: (top left) at the $N_p=0.95 \cdot 10^{12}$ and $Q'_{x,y}$=-6/-6; (top right) $N_p=0.95 \cdot 10^{12}$ and $Q'_{x,y}$=-20/-20; (bottom left) $N_p=4.3 \cdot 10^{12}$ and $Q'_{x,y}$=-4/-16; (bottom right) $N_p=4.3 \cdot 10^{12}$ and $Q'_{x,y}$=-20/-20. }
\label{FigTuneScans}
\end{figure}

The tune scans were carried out under six different conditions: i) low intensity and low chromaticity: $N_p=0.95 \cdot 10^{12}$, $Q'_{x,y}$=-6/-6; ii)  low intensity and high chromaticity: $N_p=0.95 \cdot 10^{12}$, $Q'_{x,y}$=-20/-20; iii) high intensity and medium chromaticity:  $N_p=4.3 \cdot 10^{12}$, $Q'_{x,y}$=-4/-16; iv)  high intensity and high chromaticity: $N_p=4.3 \cdot 10^{12}$, $Q'_{x,y}$=-20/-20; v) medium intensity and medium chromaticity:  $N_p=2.6 \cdot 10^{12}$, $Q'_{x,y}$=-12/-12; vi)  medium intensity and high chromaticity: $N_p=2.6 \cdot 10^{12}$, $Q'_{x,y}$=-20/-20. Note, that vertical and horizontal tunes and chromaticities varied only for the time period of 2 ms after the injection, and for the rest of the Booster cycle, they stayed as for the routine operational cycles. 

The results of the first four are presented in Fig.\ref{FigTuneScans}. One can see that, in general, an increase of either the chromaticity or intensity or both leads to reduction of the available tune space for low loss operation and generally lower optimal transmission efficiencies. Table \ref{ChromTuneOptima} summarizes the findings. 
\begin{table}[h]
\begin{center}
{\begin{tabular}{|c|c|c|c|}
\hline
$N_p,10^{12}$	& $Q'_{x,y}$=-6/-6	& $Q'_{x,y}$=-12/-12& $Q'_{x,y}$=-20/-20 \\
\hline 
0.95 &	1.5\%, 0.68/0.84 &  & 1.2\%, 0.69/0.88 \\
 & 3.5\%, 0.82/0.81 & & 2.6\%, 0.77/0.88 \\
\hline 
4.3 & 	& 3.0\%, 0.74/0.87 &  7.0\%, 0.77/0.88	\\
& & 7.0\%, 0.80/0.90 & 16\%, 0.77/0.88 \\
\hline
\end{tabular}}
\caption{Optimal working points for various injected intensities and injection chromaticities: first line in each box – the minimal intensity loss 1 ms after injection (raw B:CHG0 data, uncorrected for the B:CHG0 systematic errors and the notcher gap cleaning) and the optimal horizontal and vertical tunes $Q_x/Q_y$; the second line – same for the entire Booster cycle (at extraction).}
\label{ChromTuneOptima}
\end{center}
\end{table} 

The tune scan data reveal stronger sensitivity of the losses to the vertical tune than to the horizontal one. For example, 14-units increase of the chromaticity from -6 to -20 at $N_p=0.95 \cdot 10^{12}$ resulted in the reduction of the 90\% transmission tune area by $dQ_y$=0.05 in vertical plane while $dQ_x$=0.02 – see – see Fig.\ref{FigTuneScans} a) and b). Similarly, the change of the chromaticity from -12 to -20 for $N_p=4.3 \cdot10^{12}$ led to shrinkage of the 90\% transmission tune area by $dQ_y$=0.1 and $dQ_x$=0.05, as depicted in Fig.\ref{FigTuneScans} c) and d).  That is indicative of a stronger resonance in the vertical than the horizontal.

\section{BEAM EMITTANCE EVOLUTION}
\label{Emittance}

\subsection{Beam emittance diagnostics}
\label{EmittanceDiagnostics}

In the Booster, there are two types of instruments to measure beam sizes and therefore, transverse emittances – the multi-wires (MWs) and the ionization profile monitors (IPMs).

Vertical and horizontal MWs are installed in the extraction beam line and, therefore, can measure only the emittances of the extracted Booster beam. There are 48 wires in each instrument, spaced by 1 mm. The focusing optics function at the MW location are $\beta_x$=16.2m, $\beta_y$=25.9m and $D_x$=-1.65m. Statistical rms error of the MW emittance measurement is about 0.05 mm mrad. 

IPMs operate by collecting ions created after the ionization of residual vacuum molecules by high energy charged particle beams \cite{strehl2006beam, wittenburg2013instrum}, which are then guided to a detector by a uniform external electric field $E_{\rm ext}$. The detector consists of many thin parallel strips, whose individual signals are registered to make the beam profile signal ready for processing. Two IPMs – vertical and horizontal, are installed in the Booster at the location with $\beta_x$=6.0m, $\beta_y$=20.3m and $D_x$=1.8m \cite{zagel2010ipms}. The electric field of about 2.4 kV/cm is formed by application of $V=24$kV extracting voltage over $D=103$mm gap. The MCP based ion detector employs an array of parallel thin anode strips spaced 1.5 mm apart. IPMs are very fast and report the average rms beam sizes (determined by the Gaussian fits of the profiles) on every Booster turn. Note, that the Booster IPMs do not employ external magnetic fields to keep the trajectories of the secondaries parallel to the electric field. 

\begin{figure}[htbp]
\centering
\includegraphics[width=0.99\linewidth]{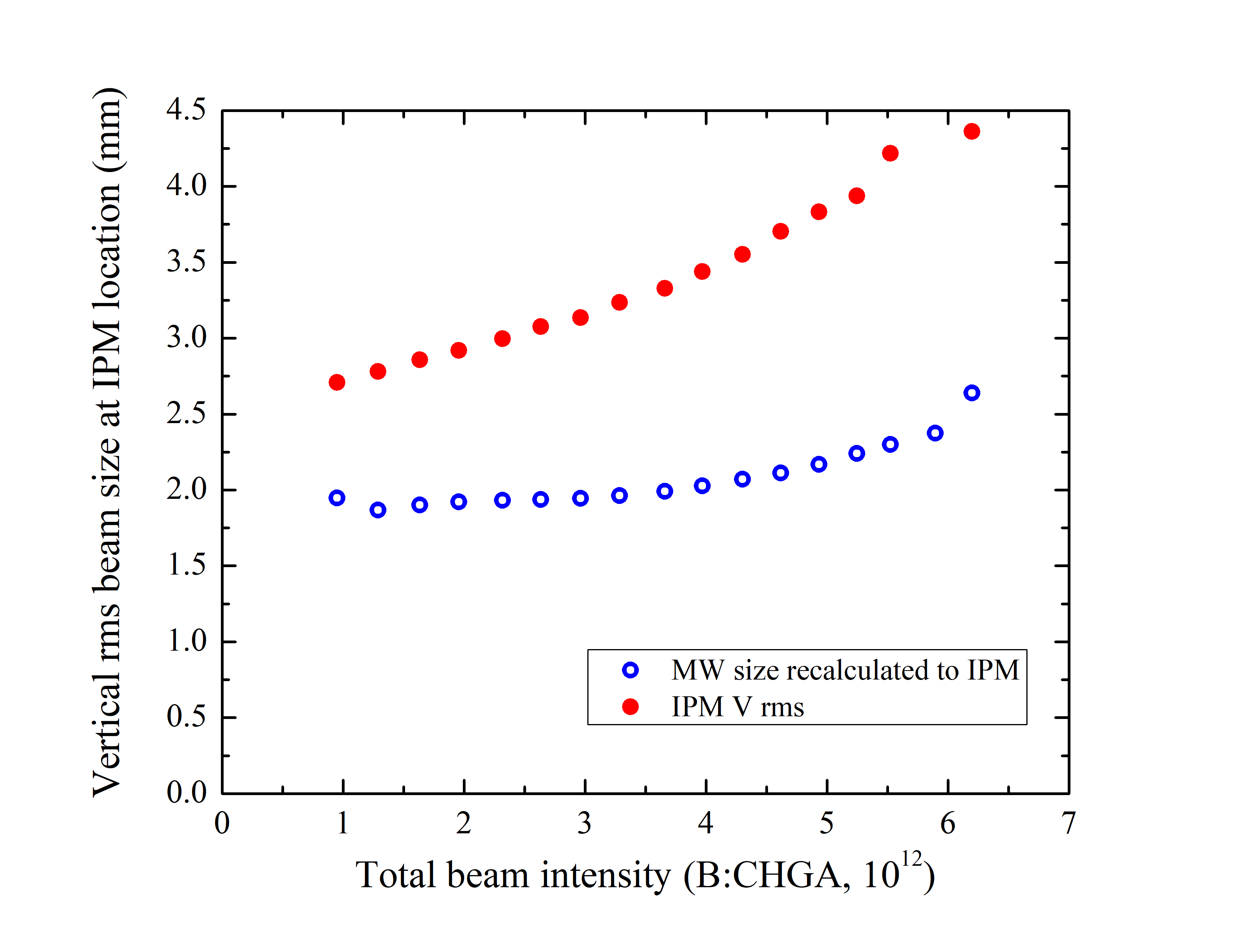}
\caption{Comparison of the measured rms IPM vertical beam sizes at extraction for different beam intensities with the rms sizes measured by the MWs and recalculated to the IPM location.}
\label{FigIPMvsMWcompare}
\end{figure}

 
Contrary to the MWs, the IPMs exhibit intrinsic dependence on the proton beam intensity as the proton space charge fields lead to transverse expansion of the cloud of ions on its way from its origin (in the proton beam) to the IPM detector plate - see Fig.\ref{FigIPMvsMWcompare}. Comprehensive theory of the IPM operation is developed in \cite{shiltsev2020ipm}. The rms size of the measured profile in the IPM $\sigma_m$ is related to the original proton beam size $\sigma_0$ as: 
\begin{equation}
\sigma_m=\sigma_0 \cdot h(N_p,\sigma_0,D,V,d) ,    
\label{EQ3}
\end{equation}
where the expansion factor $h$ can be approximated as:
\begin{equation}
h \approx 1+F\Big( \frac{2\Gamma(1/4) U_{SC} D d^{1/2}} {3 V \sigma_0^{3/2}} \Big) \big( 1+t_b/\tau_0 \big).
\label{EQ4}
\end{equation}
Here $V$ is the IPM extracting voltage (typically, 24 kV in our case) and $D$ is its HV gap (103 mm), $d$ is the distance for ions to travel from the beam orbit to the IPM collection plate, the space-charge potential for the proton beam with current $I$ is $U_{SC}$ = 30[V]$I/ \beta_p$ and gamma function $\Gamma$(1/4)$\approx$3.625.  The numerical factor $F$ is equal to 1 in the case of unbunched DC proton beam with Gaussian transverse current distribution, and $F=2\sqrt{2}/\Gamma(1/4)\approx0.78$ for uniform distribution with radius $a=2\sigma_0$. In the case of Booster, with modest expansion $h\leq2$, one can neglect minor corrections due to somewhat unequal horizontal and vertical beam sizes, but should account the correction factor $(1+t_b/\tau_0)$ due to the bunch structure of the Booster proton current. There $t_b$ is the bunch spacing (about 19 ns at the end of the Booster cycle) and characteristic time for an ion to leave the beam $\tau_0=\sqrt{2MD\sigma_0/ZeV}$, where $M$ and $eZ$ are the ion's mass and charge, that is about 22 ns for typical IPM and beam parameters at the end of the cycle. 

Also very important are intensity independent effects leading to the IPM profile smearing such as the initial velocities of the ions, finite separation between the individual IPM charge collection strips, angular misalignment of the IPM long and narrow strips with respect to the high energy proton beam orbit, etc \cite{shiltsev2020ipm}. These effects are monitor-specific, they add in quadrature and can be determined in comparison of low intensity beam sizes measured by the IPM  and by the MWs $\sigma^2_T=\sigma^2_{m,IPM} - \sigma^2_{m,MW}$ at $N_p \rightarrow 0$. For the Booster IPMs it was found that such instrumental smearing is $\sigma^2_T=2.8 \pm 0.1$ mm$^2$ and correspondingly modified Eq.(\ref{EQ3}) $\sigma^2_{IPM}=\sigma^2_T+h^2 \sigma^2_{MW}$ describes the Booster IPM data at extraction with some 5\% accuracy \cite{Shiltsev2020}.  
   

The original proton beam size $\sigma_0$ can be found from the measured and correspondingly corrected IPM value of $\sigma^*= \sqrt{\sigma^2_{m,IPM} - \sigma^2_T}$ by reversing the equation $\sigma^*=\sigma_0 (1+cN_p/\sigma_0^{3/2})$, where $c$ is deducted from  Eqs.(\ref{EQ3},\ref{EQ4}), if other parameters, such as $\sigma_T, N_p, d$ and the IPM voltage $V$ and gap $D$, are known. A simple practical algorithm gives better than $\pm$5\% approximation over the entire range of the Booster beam intensities: 
\begin{equation}
\sigma_0 \approx \frac{\sigma^*}{(1+cN_p / {\sigma^*}^{3/2})(1+\alpha c^{2N_p^2}/{\sigma^*}^2)} \, .
\label{EQ5}
\end{equation}

\begin{figure}[htbp]
\centering
\includegraphics[width=0.99\linewidth]{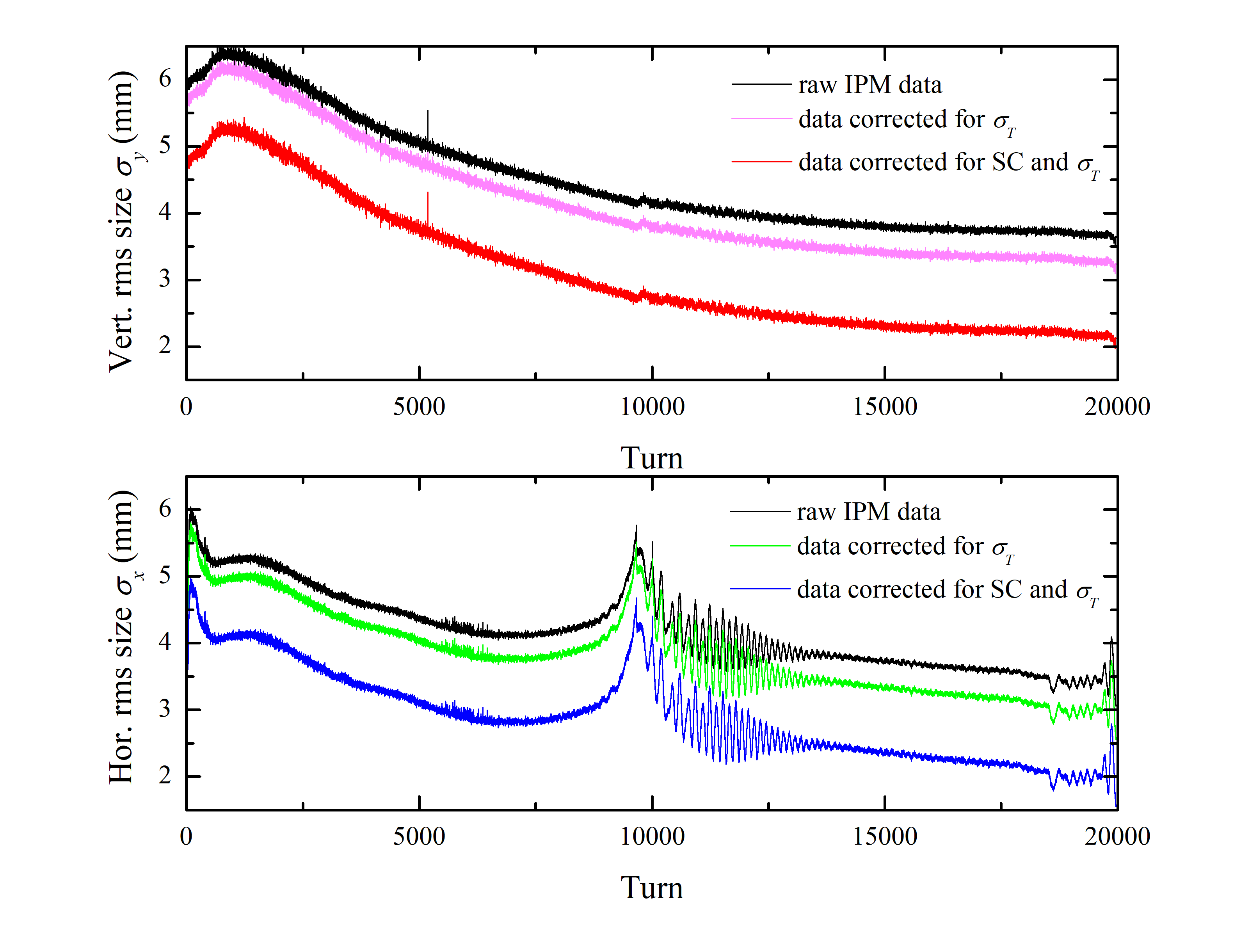}
\caption{An example of reconstruction of vertical and horizontal rms proton beam size in the 33 ms (20000 turns) acceleration cycle of the Fermilab 8 GeV Booster synchrotron with the total beam intensity of $N_p=4.6\cdot10^{12}$: time dependence of the original IPM data $\sigma^2_{m,IPM}$, the data corrected for smearing effects $\sigma^*$ and the same data after additional correction for the space-charge expansion $\sigma_{0}$ with parameters $D=103$ mm, $d=$52 mm, $V=$24 kV - see text and Eqs.(\ref{EQ4}, \ref{EQ5}).  }
\label{FigIPMcorrectionA15ATPL}
\end{figure}

For the highest beam intensity in our studies $N_p=6\cdot 10^{12}$ the factor $cN_p$=2.53 mm$^{3/2}$ and the fitting coefficient $\alpha \approx0.4$ \cite{Shiltsev2020}. Eq.(\ref{EQ5}) can now be used to find out proton beam size over the entire Booster cycle, i.e., not just for the values measured at extractions - see Fig.\ref{FigIPMcorrectionA15ATPL}. There, the black lines for the raw (uncorrected) rms vertical and horizontal beam sizes $\sigma_{m,IPM}$ as measured by the IPM at each of 20 thousand turns of the Booster acceleration cycle; the pink and green lines represent the beam sizes $\sigma^*$ corrected for the intensity independent smearing $\sigma_T$; and, finally, the true proton rms beam size $\sigma_0$ reconstructed following the above algorithm Eq.(\ref{EQ5})are represented by the red and blue lines. One can see that the overall beam size correction is about 10-15\% early in the Booster acceleration cycle when the rms beam size is about 5 mm. At the end of the cycle, with proton energy increased from 400 MeV to 8 GeV, the proton beam size is almost a factor of about 1.8 smaller than it appears in the IPM and accounting for the space-charge expansion $h(N_p,D,V,d)$ is the most important. The reconstructed beam sizes $\sigma_0$ at the end of the acceleration cycle match well the extracted beam size measured by the MWs.

\begin{figure}[htbp]
\centering
\includegraphics[width=0.99\linewidth]{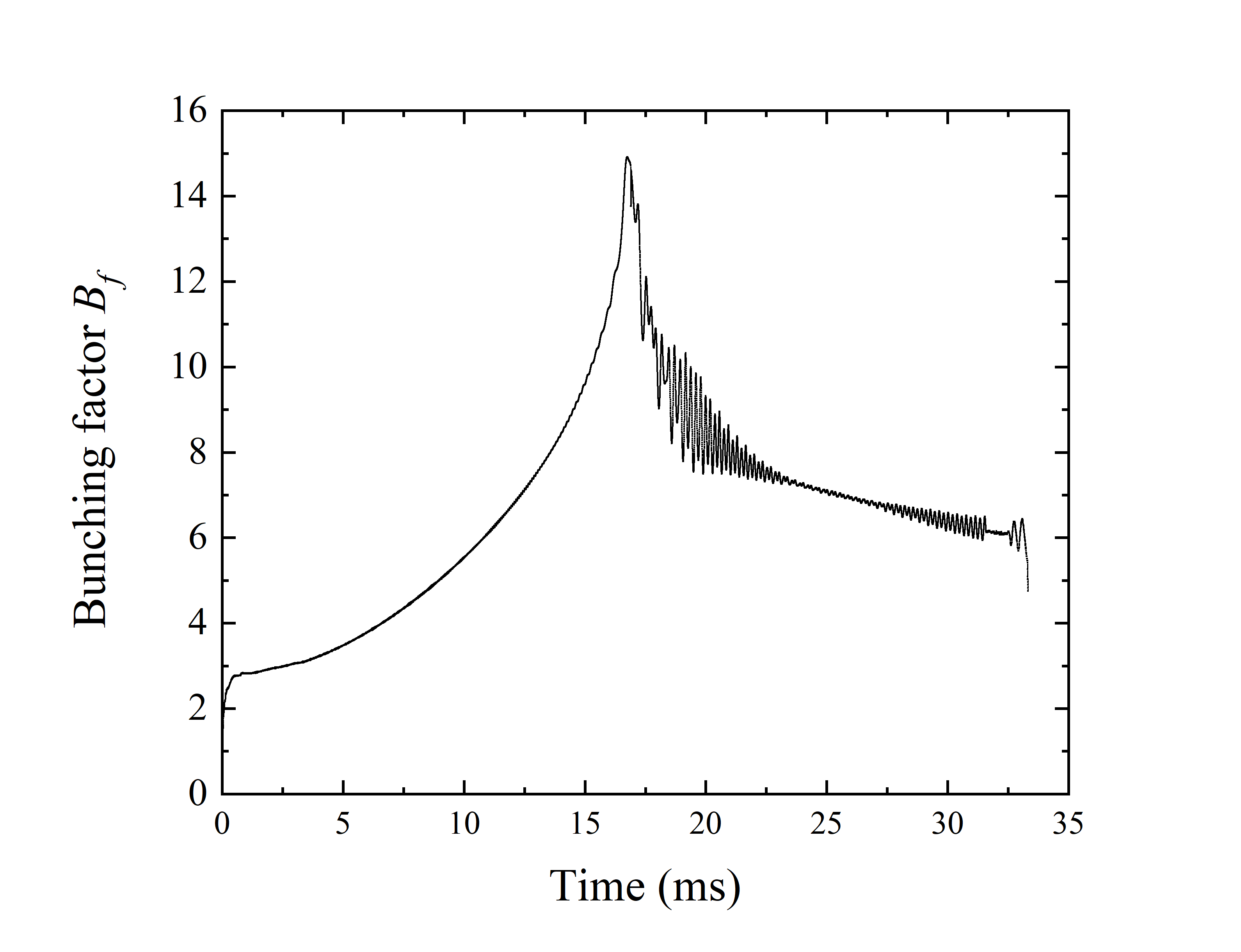}
\caption{Booster bunching factor over the accelerator ramp (inversely proportional to rms proton bunch length.)}
\label{FigBunchLength}
\end{figure}

As can be seen in Fig.\ref{FigIPMcorrectionA15ATPL}, the horizontal rms beam size exhibits significant oscillations with twice the synchrotron frequency after the transition. Such oscillations arise from the mismatch between longitudinal focusing of bunch fields before and after transition and the effect gets bigger with intensity. 
Horizontal IPM is located at small beta-function and high dispersion, so, compared to the betatron size, the dispersive contribution is large $D_x (\delta p/ p) \ge \varepsilon_x \beta_x/(\beta_p \gamma_p)$ and variations in the momentum spread $(\delta p/p)$ result in about $\pm20$\% oscillations in $\sigma_x$. Booster is well decoupled and, consequently,  the vertical dispersion is small. Therefore a little perturbation observed at transition is most probably due to reaction of the the IPM profile expansion factor $h$ to variation of the bunching factor $B_f=\sqrt{2\pi}/\sigma_s$ - the ratio of the peak to average proton beam current - which peaks at the transition - see Fig.\ref{FigBunchLength}. The effect is small but becomes more pronounced at higher intensities (see Refs.\cite{shiltsev2020ipm}  below).  

\subsection{Beam emittance vs intensity}
\label{EmittanceInstensity}

The injected Linac $H^-$ beam comes with the rms normalized transverse emittances $\varepsilon \sim 1\pm 0.2$ $\pi$ mm mrad \cite{bhat2015injemm}. Due to small injection errors, optics mismatch and multiple scattering in the foil the initial emittance of the proton beam circulating in the Booster gets to about  $\sim 1.2\pm 0.2$ $\pi$ mm mrad. The latter effect correlates with the total beam intensity. Indeed, in the process of multi-turn charge exchange injection, each passage of the stripping foil leads to the emittance growth of: 
\begin{equation}
\Delta \varepsilon_{x,y} \approx \beta_p\gamma_p {\beta_{x,y} \over 2}{l \over X_0} \Big( \frac{13.6 \, \rm{MeV}}{\beta_p pc} \Big)^2 [1+0.0038 \ln{(l/X_0)}]\, .
\label{EQ6}
\end{equation}
For the Booster carbon foil thickness $l$=380 $\mu$g/cm$^2$, radiation length $X_0$=42.7 g/cm$^2$ and momentum $p$=953 MeV/c that gives 0.0032 $\pi$ mm mrad per turn in the horizontal plane and 0.011 $\pi$ mm mrad per turn in the vertical plane. During the injection the beam moves across and off the foil, so the effective number of turns is about ($N_{turns}$+29)/2 turns \cite{eldred2019foil}. Given that the total circulating beam intensity $N_p$ scales linearly with $N_{turns}$, the estimated emittance increase at the end of injection grows with the intensity and for the nominal $N_{turns}$=14 turns injection  $\Delta \varepsilon_{y, foil} \approx 0.24 \pi$ mm mrad and  $\Delta \varepsilon_{x, foil} \approx 0.07 \pi$ mm mrad.
\begin{figure}[htbp]
\centering
\includegraphics[width=0.99\linewidth]{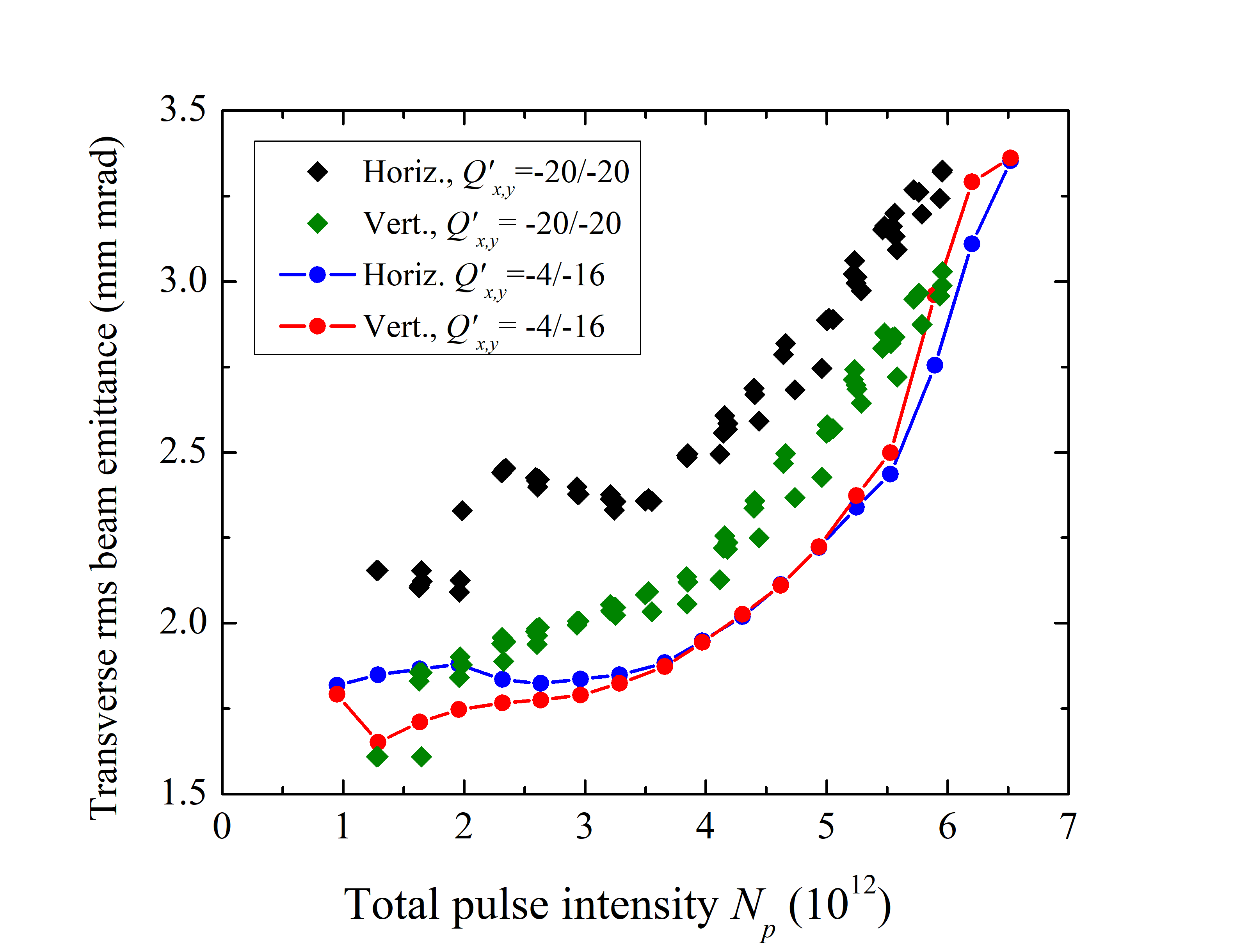}
\caption{Booster beam emittance measured by MWs at extraction vs the total proton intensity.}
\label{FigEmmMWs}
\end{figure}

Measured Booster beam emittances at extraction exhibit strong dependence on the total proton intensity $N_p$ as shown in Fig.\ref{FigEmmMWs}. The MWs data taken at the nominal operational chromaticities at injection $Q’_{x,y}$=-4/-16 show strong dependence on the beam intensity and for both planes can be approximated as: 
\begin{equation}
\varepsilon_{extr}\mathrm{[\pi \, mm \, mrad]} \approx 1.7 + 2.1 \cdot \Big( { N_p \over {7 \cdot 10^{12}}} \Big)^{4\pm0.3} \, ,
\label{EQ7}
\end{equation}


The emittance growth is strongly dependent on the chromaticity, too, and at the nominal intensity $N_p= 4.5\cdot 10^{12}$ the emittance increases from about 2.1 $\pi$ mm mrad to some 2.4 $\pi$ mm mrad (vertical) and 2.7 $\pi$ mm mrad (horizontal) if the operational chromaticity at the first ms after the injection is changed from $Q’_{x,y}$=-4/-16 to -20/-20. The beam emittance increase with intensity and chromaticity strongly correlates with the intensity losses - see Eqs.(\ref{EQ1}, \ref{EQ2}) and Figs.\ref{FigBoosterFractLoss}, \ref{FigChromLoss}.  

The IPM rms beam sizes $\sigma_{0,x,y}(t)$ measured over the Booster acceleration cycle from injection to extraction and properly corrected following the analysis of preceding Sec.\ref{EmittanceDiagnostics} and Eq.(\ref{EQ5}) can be used for the emittance calculations $\varepsilon_{y}=(\beta_p \gamma_p)\sigma^2_{0,y} /\beta_y$, $\varepsilon_{x}=(\beta_p \gamma_p)(\sigma^2_{0,y}-D^2_x (\delta p/p)^2) / \beta_x$, where $\delta p/p$ is the rms energy spread. Of course, all the factors are now time dependant: the relativistic factors $\beta_p(t)$ and $\gamma_p(t)$ are well known - see, e.g. Fig.\ref{FigBoosterRamp}; the beta-functions $\beta_{x,y}$ at the IPM locations vary in the cycle within $\sim$10\%. Calculations of Booster emittance take all these effects into account, they agree with the MWs data at extraction to within 10\% and exhibit no emittance growth at the lowest intensities - see \cite{Shiltsev2020}. 


\begin{figure}[htbp]
\centering
\includegraphics[width=0.99\linewidth]{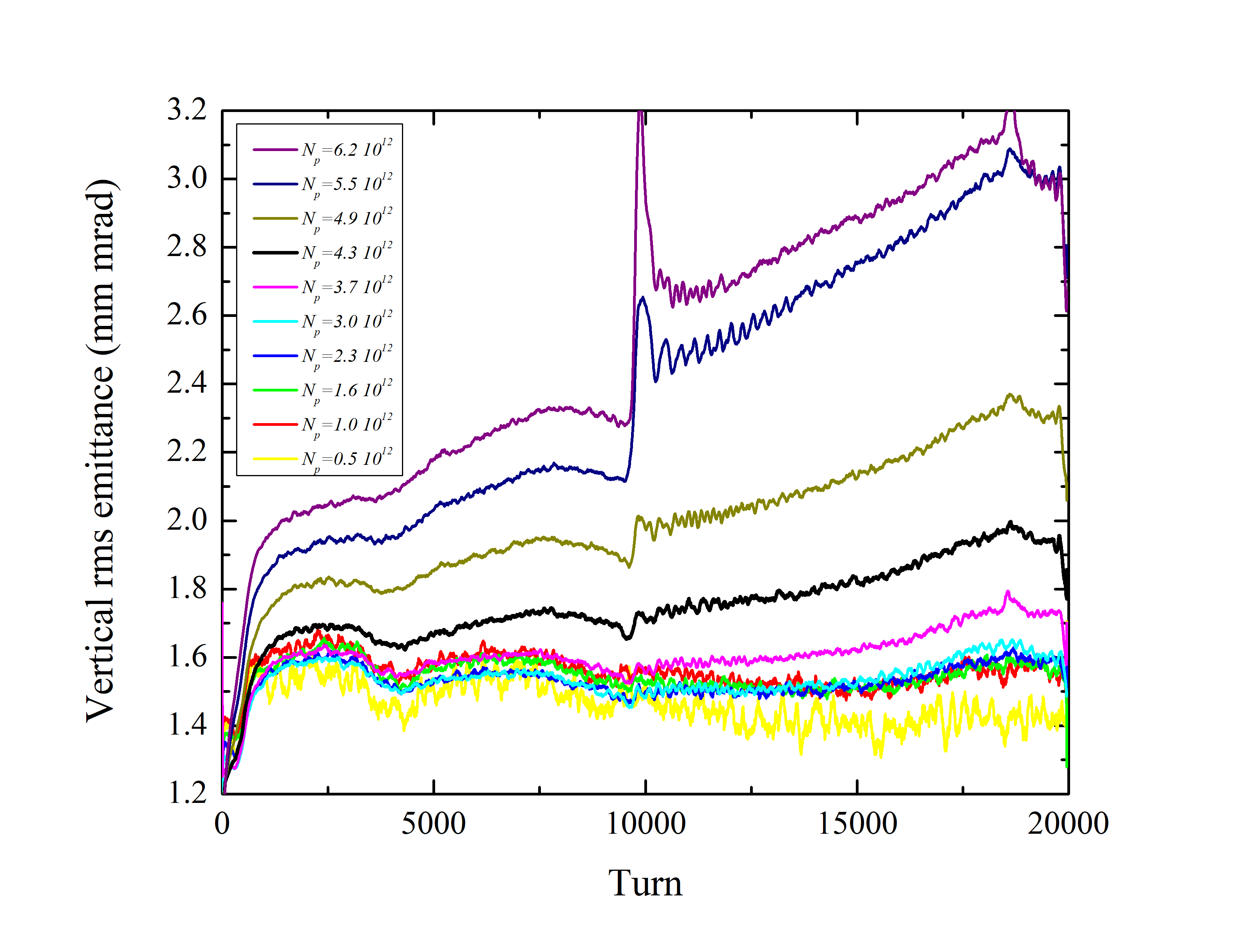}
\caption{Evolution of the IPM vertical emittance in the Booster cycle at different intensities $N_p$ from 0.5$\cdot 10^{12}$ (2 turns injection) to 6.2$\cdot 10^{12}$ (20 turns injection). All the data are smoothed by a 100 turn running window averaging. }
\label{FigVertEmmvsNP}
\end{figure}   

The resulting Booster IPM vertical beam emittance evolution over the acceleration cycle is shown in Fig.\ref{FigVertEmmvsNP} for a wide range of intensities $N_p$ from 0.5$\cdot 10^{12}$ (2 turns injection) to 6.2$\cdot 10^{12}$ (20 turns). 
For example, the emittance $\varepsilon_y(t)$ at the operational intensity of $N_p=4.3 \cdot 10^{12}$ is shown in black line in Fig.\ref{FigVertEmmvsNP}. Shortly after injection is it about 1.3$\pm$0.1 mm mrad and evolves to 1.9$\pm$0.1 mm mrad at extraction – in an good agreement with the MW emittance data shown in Fig.\ref{FigEmmMWs}. In general, one can see that up to about $N_p=3.7 \cdot 10^{12}$ (12 turns injection, pink line) the emittance is not growing much in the cycle and is about  1.4-1.6 mm mrad. Above that intensity the emittance evolution exhibits several features: i) fast growth over the first 2000-3000 turns, ii) steady growth for the rest of the cycle; iii) spikes at the time of transition and minor oscillations afterwards; and iv) 5-10\% variations at the end of the cycle. The last two effects are presumably instrumental. Significant variation of the bunching factor at the transition shown in Fig.\ref{FigBunchLength} does affect the IPM profile expansion $h$ – see Eq.(\ref{EQ4}). At the end of the acceleration cycle, the proton beam position in the IPMs varies over the last 2000 turns by as much as 6 mm in the horizontal plane, thus, affecting the IPM profile expansion factor $h_{y}$ in the vertical plane that scales with the distance from the beam orbit to the IPM collection plate as $d_{x, y}^{1/2}$ - see Eq.(\ref{EQ4}), while bunch rotation in longitudinal phase-space prior the extraction at the very last hundreds of turns results in a smaller momentum spread and longer bunches - again, leading to minor variation of the IPM expansion factor $h$. 

\begin{figure}[htbp]
\centering
\includegraphics[width=0.99\linewidth]{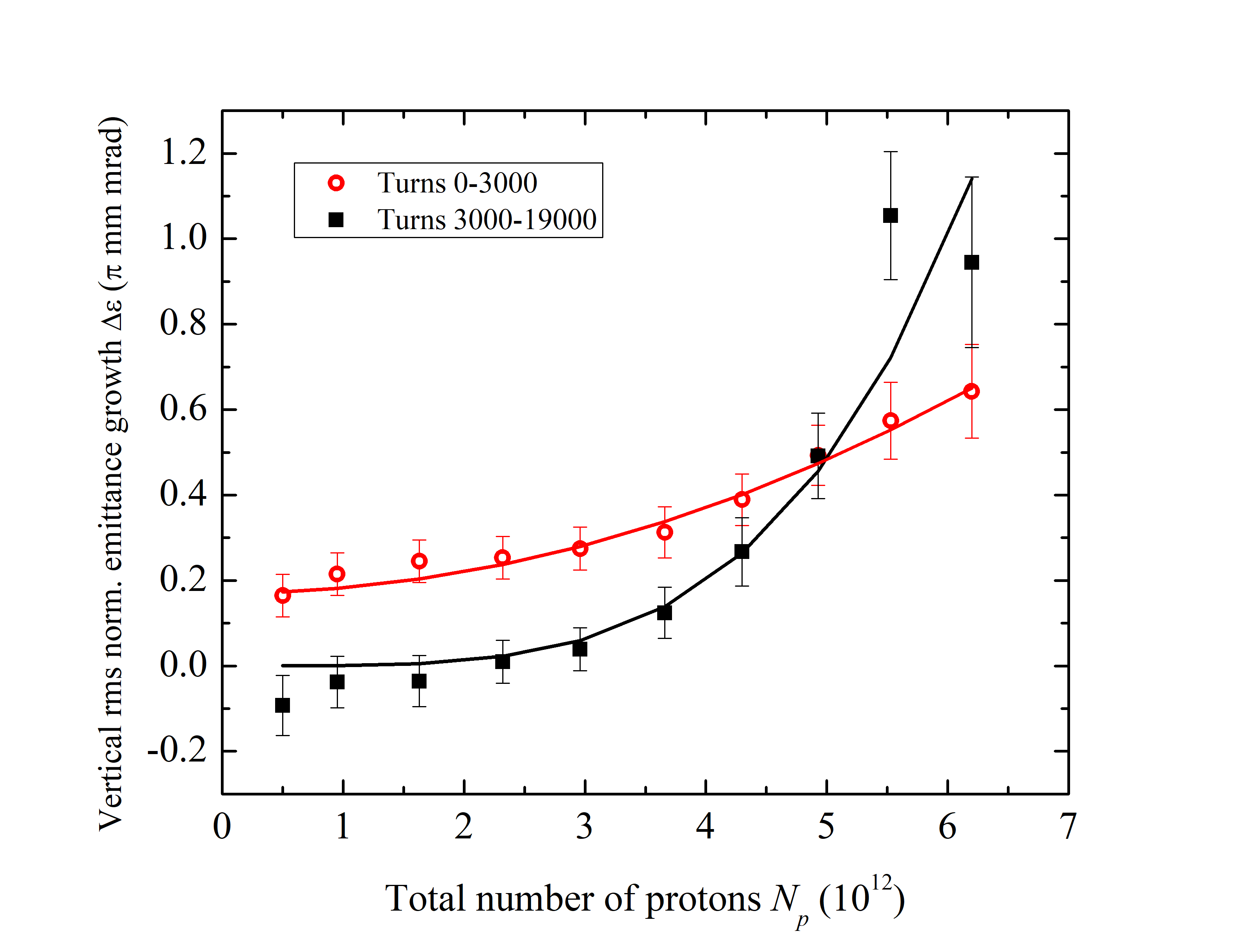}
\caption{Vertical rms emittance growth vs Np : (red circles) over the first 3000 turns; (black squares) from 3000 to 19000 turns. The data points are calculated for the IPM emittance values averaged over five hundred turns 0-500, 3000-3500, 19000-19500. The error bars indicate estimated statistical uncertainty. Red and black solid line are for the approximations Eq.(\ref{EQ8}) and (\ref{EQ9}), respectively.
}
\label{FigdEmm3000}
\end{figure}

The most dominant are the first two effects. Fig.\ref{FigdEmm3000} shows how they depend on the beam intensity. The fast rms vertical emittance growth over the first 3000 turns is most probably due to record strong proton space-charge effects (see below) and scales approximately as: 
\begin{equation}
\Delta \varepsilon_{y, 3000}\mathrm{[\pi \, mm \, mrad]} \approx 0.17 + 0.61 \cdot \Big( {N_p \over {7 \cdot 10^{12}}} \Big)^{2} \, .
\label{EQ8}
\end{equation}

Slow emittance increase is roughly linear in time over the next 16000 turns and gets as big as 1 mm mrad, or 30\% of the emittance, at $N_p=6.2\cdot 10^{12}$ and can be approximated as :
\begin{equation}
\Delta \varepsilon_{y, 3000-19000}\mathrm{[\pi \, mm \, mrad]} \approx 1.85 \cdot \Big( { N_p \over {7 \cdot 10^{12}}} \Big)^{4} \, .
\label{EQ9}
\end{equation}
The nature of that effect might be related to the multipacting of electrons in the beam and/or beam-induced vacuum activity but that still needs to be confirmed \cite{eldred2020ecloudstudies}.

\subsection{Space-charge tune shift}
\label{DiscussdQSC}

Space-charge tuneshift parameter $\Delta Q_{SC}$ is a commonly used figure of merit for beam dynamics. It is equal to  \cite{wei2003synchrotrons}: 
\begin{equation}
\Delta Q_{SC} = \frac{N_p r_p B_f} {4\pi \varepsilon \beta_p \gamma_p^2} R \, ,
\label{EQ10}
\end{equation}
where $N_p$ is the total intensity assuming that the bunches fill all RF buckets, $r_p$ is the classical proton radius, $B_f$ is the bunching factor (the ratio of the peak to average bunch current),  $\varepsilon$ is the normalized rms beam emittance, $\beta_p$ and $\gamma_p$ are relativistic Lorentz factors. Factor $R\lesssim 1$ accounts for unequal average beam size ratio around the ring and, e.g., for the vertical plane it is equal to $\big< 2/(1+\sigma_x/\sigma_y)\big>$. The tuneshift is negative but we omit the minus sign for simplicity. In operational circular rapid cycling accelerators, the space-charge parameter usually does not exceed 0.2-0.4 to avoid beam losses \cite{weiren1999icfa, wei2003synchrotrons, tang2013rcs}. 

\begin{figure}[htbp]
\centering
\includegraphics[width=0.99\linewidth]{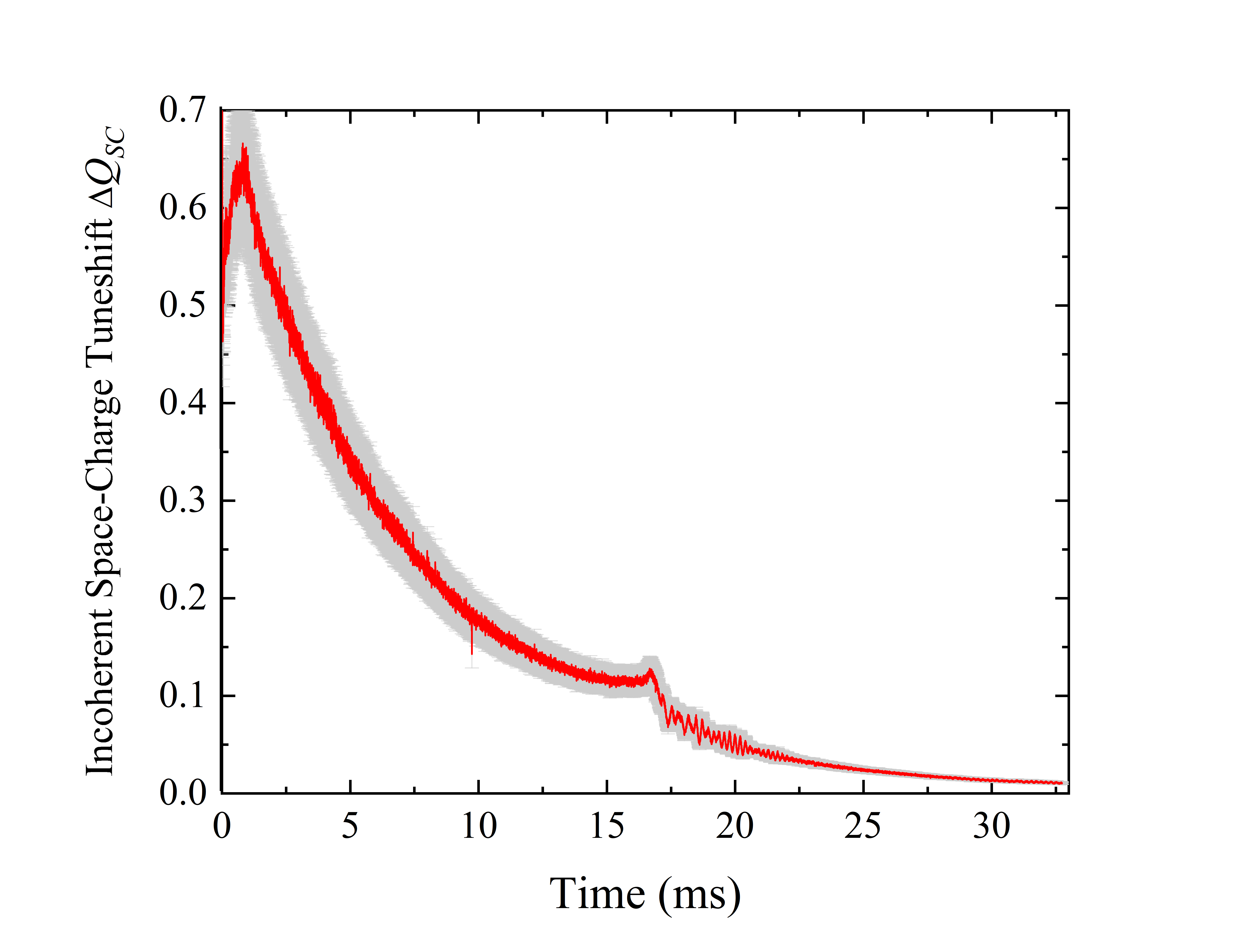}
\caption{Calculated vertical space-charge tuneshift parameter for the Booster cycle with $N_p=4.6\cdot 10^{12}$. (Shaded area indicates 10\% uncertainty, mostly due to the IPM emittance calculations). 
}
\label{FigVdQSC}
\end{figure}

Fig.\ref{FigVdQSC} shows the vertical SC tuneshift parameter $\Delta Q_{SC}(t)$  calculated for the Fermilab Booster acceleration cycle on base of measured $N_p(t), \varepsilon_y(t), B_f(t)$ and known $\beta_p(t)$ and $\gamma_p(t)$. One can see that the calculated space-charge parameter quickly grows after the injection due to fast bunching early in the acceleration cycle, then falls down due to acceleration and emittance increase, and exhibits some temporary increase at the transition before ending at $\sim$0.01 prior to extraction. The maximum Booster tuneshift parameter peaks at about 1 ms after injection to $\Delta Q_{SC} \simeq 0.65$ and stays above 0.3 until about 6 ms ($\sim$3000 turns). Naturally, the corresponding incoherent space-charge tune spread does not easily fit the available tune space between most dangerous resonances, such as half-integer ones, and that results in strong resonant excitation of the proton dynamics and eventual particle losses at the machine aperture. At the highest beam intensity studied $N_p=6.2\cdot 10^{12}$ the maximum space-charge tuneshift parameter $\Delta Q_{SC}$ peaks at $\simeq 0.75$.

\section{SUMMARY, DISCUSSION}
\label{Discussion}

The phenomena of the Booster beam intensity losses, transverse emittance growth and longitudinal beam excitation at the transition are found to be very much intensity dependent and strongly interconnected. Our experimental studies indicate that the losses that occur early in the cycle are most probably due to the space-charge driven vertical beam size expansion that leads to the flux of protons on the collimator-limited machine aperture with characteristic acceptance $A_y\simeq 20-25 \, \mathrm{[\pi \, mm \, mrad]}$\cite{Shiltsev2020}. Both the fractional beam loss Eqs.(\ref{EQ1}, \ref{EQ2}) and the emittance growth Eq.(\ref{EQ8}) highly nonlinearly depend on the total circulating beam intensity $N_p$. 

Beam losses at and after the transition crossing while small under conditions of optimal operational tune up the nominal intensity $N_p=4.6\cdot 10^{12}$ exhibit a threshold-like behavior  above $6\cdot 10^{12}$. IPM measurements indicate small vertical beam size at the transition while the horizontal size, dominated by the dispersive contribution $D_x(\delta p/p)$, becomes very large, so it is natural to assume that the proton losses at the transition end up at the horizontal aperture. 

Booster is a lynch pin of the Fermilab proton accelerator complex and its transmission efficiency has always been subject of continuous monitoring and several studies for almost four decades - see, e.g. \cite{moore1981dependence, popovic1998performance, chou2003fermilab}. Naturally, the Booster performance at present significantly exceeds that of the past - compare current transmission $\sim$95\% at $N_p=(4.5-5)\cdot 10^{12}$ with that of $\sim$(70-75)\% at $N_p=(3.5-4)\cdot 10^{12}$ in 1980s and 1990s. What makes this work distinct is that for the first time we have performed a simultaneous comprehensive analysis of the proton loss dependence on the total beam intensity, machine tunes and chromaticities, and, also, performed a cross-calibration of the toroid intensity monitor and the beam loss monitors. 

Nonlinear dependence of the extracted Booster beam emittance on $N_p$  was reported in many previous measurements with MultiWire profile monitors - see, e.g.\cite{moore1981dependence, ankenbrandt1987limits, holmes1997maininj}. The Booster Ionization Profile Monitors operating in the ion collection mode without external magnetic field are known to be extremely valuable tools for fast beam size diagnostics of the circulating beam during acceleration \cite{graves1995ipm, zagel2010ipms, amundson2003calibration}. At the same time, strong space-charge forces of the high intensity proton beam lead to significant, factor of 2 or more, expansion of the rms beam size reported by the IPMs compared to the original proton beam size. To be certain that we properly account that and other, intensity-independent, effects, we - in addition to following the theoretical recipes of Ref. \cite{shiltsev2020ipm} as described in Sec.\ref{EmittanceDiagnostics} - has performed a systematic calibration of the IPM with the MultiWires measurements. As the result, we have achieved $O$(10)\% accuracy in the beam emittance reconstruction using the measured IPM rms beam sizes, experimentally determined intensity-independent instrumental dispersion $\sigma_T$, known beam intensity $N_p$ and the IPM parameters such as extracting electric field $E_{\rm ext}=V_0/D$ and the distance $d$ from the beam orbit to the IPM detector. 

Our data indicate existence of two phenomena leading to the proton beam emittance growth: i) the space-charge driven expansion over the first few thousand turns; and ii) steady emittance increase over the rest of the acceleration cycle. Both effects grow faster than linear with the proton beam intensity, resulting in the final (extracted) beam emittance having a significant component $\propto \, N_p^4$ - see Eq.(\ref{EQ7}), which also dependent on the machine chromaticity at the injection energy. An attempt of a similar analysis of the Booster IPM profiles has been undertaken in \cite{huang2006emittance}. It was based on the phenomenological approximations for the IPM profile space-charge induced expansion effects developed in \cite{amundson2003calibration}, and reported space-charge driven emittance growth early in the cycle was found to be a) significantly, 3-4 times bigger than reported here; and b) scaling approximately linear with intensity $\propto N_p$. These observations also show improbably large normalized rms emittances of high intensity beams prior to extraction $\sim 10\mathrm{[\pi \, mm \, mrad]}$, thus, directly contradicting our MultiWires measurements and what we know about the Booster acceptance - that make us to suspect a systematic error in the IPM data analysis. Notably, the steady emittance increase during the acceleration cycle was observed, too, though again, more than twice of we report here in Eq.(\ref{EQ9}). 

Our experimental studies of the Booster losses and emittance evolution augment similar investigations at other high-intensity proton synchrotrons, see, e.g., \cite{franchetti2003space, franchetti2010experiment, asvesta2020identification, molodozhentsev2007space, ohmi2014study, hotchi2017achievement}. They are of great importance to predict the machine operational conditions in the era of upcoming new 800 MeV injector (PIP-II) and have to be continued. These studies and foreseen future operation with up to 50\% higher beam intensities would greatly benefit from improved accuracy of the fast beam intensity diagnostics and of the IPM emittance analysis. The losses at transition are believed to result from complicated 3D dynamics including transverse and longitudinal coherent instabilities. Future investigations of the dependence on the RF voltage, chromaticities, tunes and other machine parameters will help to better identify the transition loss origin. ESME~\cite{bhat2020esme} and BLonD~\cite{derwent2020blond} simulations are also underway to study the impact of $\gamma_t$ jumps on transition in the PIP-II era.

Strong intensity dependence of the space-charge induced losses in the first few thousands turns after the injection is also of concern.  
Indeed, let us consider an RCS accelerator, like the Fermilab Booster, operationally limited by the uncontrolled radiation level in accelerator enclosures at $W=f_0 \int E_k dN_p$ (here $f_0$ is the cycle rate and $E_k$ is the kinetic energy of the lost particle) - typically it is about 1 W per a meter of machine circumference. Under such limit, the tolerable fractional beam intensity loss is 
\begin{equation}
{\Delta N_p \over N_p} = \frac{W} {(1-\eta) N_p E_k f_0} \, ,
\label{EQ11}
\end{equation}
where $\eta$ is the efficiency of the collimation system that directs the losses into dedicated beam absorbers or dumps. Obviously, the losses should get smaller with the increase of beam intensity, energy and power. On the contrary, many beam physics phenomena, such as, e.g., repelling forces of the proton beam's own space-charge lead to increase of beam sizes and particle losses at higher beam intensities. Let us, following Eq.(\ref{EQ1}), consider the case when the space-charge induced losses grow with the space-charge tuneshift $\Delta Q_{SC}$ as:
\begin{equation}
{\Delta N_p \over N_p} \propto \alpha \Delta Q_{SC}^\kappa  ,
\label{EQ12}
\end{equation} 
where $\alpha$ is a machine dependent constant and the exponent $\kappa$ is about 3. Then one gets from Eqs.(\ref{EQ11}, \ref{EQ13}) the maximum operational intensity within the loss power limit $W$ :
\begin{equation}
N_p^{max} \propto \Big( {W \over 1-\eta} \Big)^{1 \over \kappa+1} \Big( {\varepsilon \over B_f} \Big)^{\kappa \over \kappa+1} \frac{\gamma_p^{{2 \kappa -1/2} \over \kappa+1}} {\big( \alpha f_0 \big)^{1 \over \kappa+1}}   \, \, .
\label{EQ13}
\end{equation} 
From that, one can see that there are several paths to the increase of the maximum intensity \cite{ShiltsevMPLA} but none leads to direct one-to-one increase of $N_p^{max}$. For the case of the PIP-II Booster and assuming $\kappa=3$, Eq.(\ref{EQ13}) anticipates that a 3-fold improvement in collimation system efficiency (e.g. from $\eta=0.55$ to 0.85)~\cite{kapin2017collimation} leads to a 3$^{1/4}$ increase in $N_p^{max}$, or 32\%. For the two-fold increase in Booster injection energy $E_k$ from 400 MeV to 800 MeV with a new PIP-II linac~\cite{PIPII2017lebedev}, Eq.(\ref{EQ13}) predicts that one should be able to safely increase $N_p^{max}$ by 41\% from the current operational value, except the increase in cycle rate $f_0$ from 15 Hz to 20 Hz under the same plan will cut the expected benefit to just 31\%. Between the two improvements (neglecting any beneficial effects of PIP-II injection painting), Eq.(\ref{EQ13}) at $\kappa=3$ favors the 44\% increase in intensity from 4.5$\cdot 10^{12}$ to 6.5$\cdot 10^{12}$ with the proposed improvements for the PIP-II Booster.

More broadly, Eq.(\ref{EQ13}) evaluates strategies for improving intensity in other high-intensity rings. Flattening the longitudinal bunch current profile, e.g. by using additional 2nd or 3rd harmonics RF systems, leads to reduction of the bunching factor and the factor of two smaller $B_f$ could lead to 1.68 times higher maximum intensity. Acceleration of twice larger emittance beams would give about the same effect, but it usually not possible within available machine aperture of existing machines. At ultimate intensities, significant promise in loss reduction lies in improved beam dynamics that would make $\alpha$ and $\kappa$ smaller, for example by injection “painting” to make the space-charge forces more uniform, by compensation of the most detrimental resonant driving terms (including enforcement perfect periodicity in machine focusing optics), via the space-charge compensation using electron lenses \cite{shiltsev2015electron}, or by implementation of the non-linear integrable optics \cite{danilov2010nonlinear}. For the latter two topics in particular, an R\&D program is underway at the Fermilab IOTA facility \cite{antipov2017iota}. 

\section*{Acknowledgements} 
We would like to thank C.Y. Tan, C. Bhat, Yu. Alexahin, A. Burov, S. Nagaitsev, W. Pellico and R. Thurman-Keup for numerous discussions on the topics of this study and S. Chaurize, V. Kapin and K. Triplett for their invaluable help with experimental Booster beam studies. In addition, the Summer 2019 Booster beam study campaign involved N. Eddy, C. Jensen, J. Larson, and H. Pfeffer of  Fermilab,  H. Bartosik, N. Biancacci, M. Carla, A. Saa Hernandez, A. Huschauer, F. Schmidt of CERN, D. Bruhwiler, J. Edelen of the Radiasoft SBIR company and V. Kornilov of GSI. We greatly appreciate their fruitful cooperation and the spirit of international beam physics collaboration.  

 Fermilab is supported by U.S. Department of Energy, Office of Science, Office of High Energy Physics, under Contract No. DE-AC02-07CH11359.
  

\bibliography{Booster_SC.bib}

\begin{thebibliography}{70}%
\makeatletter
\providecommand \@ifxundefined [1]{%
 \@ifx{#1\undefined}
}%
\providecommand \@ifnum [1]{%
 \ifnum #1\expandafter \@firstoftwo
 \else \expandafter \@secondoftwo
 \fi
}%
\providecommand \@ifx [1]{%
 \ifx #1\expandafter \@firstoftwo
 \else \expandafter \@secondoftwo
 \fi
}%
\providecommand \natexlab [1]{#1}%
\providecommand \enquote  [1]{``#1''}%
\providecommand \bibnamefont  [1]{#1}%
\providecommand \bibfnamefont [1]{#1}%
\providecommand \citenamefont [1]{#1}%
\providecommand \href@noop [0]{\@secondoftwo}%
\providecommand \href [0]{\begingroup \@sanitize@url \@href}%
\providecommand \@href[1]{\@@startlink{#1}\@@href}%
\providecommand \@@href[1]{\endgroup#1\@@endlink}%
\providecommand \@sanitize@url [0]{\catcode `\\12\catcode `\$12\catcode
  `\&12\catcode `\#12\catcode `\^12\catcode `\_12\catcode `\%12\relax}%
\providecommand \@@startlink[1]{}%
\providecommand \@@endlink[0]{}%
\providecommand \url  [0]{\begingroup\@sanitize@url \@url }%
\providecommand \@url [1]{\endgroup\@href {#1}{\urlprefix }}%
\providecommand \urlprefix  [0]{URL }%
\providecommand \Eprint [0]{\href }%
\providecommand \doibase [0]{https://doi.org/}%
\providecommand \selectlanguage [0]{\@gobble}%
\providecommand \bibinfo  [0]{\@secondoftwo}%
\providecommand \bibfield  [0]{\@secondoftwo}%
\providecommand \translation [1]{[#1]}%
\providecommand \BibitemOpen [0]{}%
\providecommand \bibitemStop [0]{}%
\providecommand \bibitemNoStop [0]{.\EOS\space}%
\providecommand \EOS [0]{\spacefactor3000\relax}%
\providecommand \BibitemShut  [1]{\csname bibitem#1\endcsname}%
\let\auto@bib@innerbib\@empty
\bibitem [{\citenamefont {Shiltsev}(2011)}]{shiltsev2011complex}%
  \BibitemOpen
  \bibfield  {author} {\bibinfo {author} {\bibfnamefont {V.}~\bibnamefont
  {Shiltsev}},\ }\bibfield  {title} {\bibinfo {title} {On performance of high
  energy particle colliders and other complex scientific systems},\ }\href@noop
  {} {\bibfield  {journal} {\bibinfo  {journal} {Modern Physics Letters A}\
  }\textbf {\bibinfo {volume} {26}},\ \bibinfo {pages} {761} (\bibinfo {year}
  {2011})}\BibitemShut {NoStop}%
\bibitem [{\citenamefont
  {Shiltsev}(2020{\natexlab{a}})}]{shiltsev2020PhysToday}%
  \BibitemOpen
  \bibfield  {author} {\bibinfo {author} {\bibfnamefont {V.}~\bibnamefont
  {Shiltsev}},\ }\bibfield  {title} {\bibinfo {title} {Particle beams behind
  physics discoveries},\ }\href@noop {} {\bibfield  {journal} {\bibinfo
  {journal} {Physics Today}\ }\textbf {\bibinfo {volume} {73}},\ \bibinfo
  {pages} {32} (\bibinfo {year} {2020}{\natexlab{a}})}\BibitemShut {NoStop}%
\bibitem [{\citenamefont
  {Shiltsev}(2020{\natexlab{b}})}]{shiltsev2020superbeams}%
  \BibitemOpen
  \bibfield  {author} {\bibinfo {author} {\bibfnamefont {V.}~\bibnamefont
  {Shiltsev}},\ }\bibfield  {title} {\bibinfo {title} {Superbeams and neutrino
  factories—two paths to intense accelerator-based neutrino beams},\
  }\href@noop {} {\bibfield  {journal} {\bibinfo  {journal} {Modern Physics
  Letters A}\ ,\ \bibinfo {pages} {2030005}} (\bibinfo {year}
  {2020}{\natexlab{b}})}\BibitemShut {NoStop}%
\bibitem [{\citenamefont {Wei}(2003)}]{wei2003synchrotrons}%
  \BibitemOpen
  \bibfield  {author} {\bibinfo {author} {\bibfnamefont {J.}~\bibnamefont
  {Wei}},\ }\bibfield  {title} {\bibinfo {title} {Synchrotrons and accumulators
  for high-intensity proton beams},\ }\href@noop {} {\bibfield  {journal}
  {\bibinfo  {journal} {Reviews of Modern Physics}\ }\textbf {\bibinfo {volume}
  {75}},\ \bibinfo {pages} {1383} (\bibinfo {year} {2003})}\BibitemShut
  {NoStop}%
\bibitem [{\citenamefont {Zwaska}\ \emph {et~al.}(2018)\citenamefont {Zwaska}
  \emph {et~al.}}]{zwaska2018targets}%
  \BibitemOpen
  \bibfield  {author} {\bibinfo {author} {\bibfnamefont {R.}~\bibnamefont
  {Zwaska}} \emph {et~al.},\ }\bibfield  {title} {\bibinfo {title} {Multi{-MW}
  targets for next-generation accelerators},\ }in\ \href@noop {} {\emph
  {\bibinfo {booktitle} {Proc. of International Particle Accelerator Conference
  (IPAC'18, April 29 - May 4, 2018, Vancouver, BC, Canada)}}}\ (\bibinfo
  {organization} {JACoW},\ \bibinfo {year} {2018})\BibitemShut {NoStop}%
\bibitem [{\citenamefont {Simos}\ \emph {et~al.}(2019)\citenamefont {Simos},
  \citenamefont {Hurh}, \citenamefont {Dooryhee}, \citenamefont {Snead},
  \citenamefont {Sprouster}, \citenamefont {Zhong}, \citenamefont {Zhong},
  \citenamefont {Ghose}, \citenamefont {Kotsina}, \citenamefont {Ammigan} \emph
  {et~al.}}]{simos2019target}%
  \BibitemOpen
  \bibfield  {author} {\bibinfo {author} {\bibfnamefont {N.}~\bibnamefont
  {Simos}}, \bibinfo {author} {\bibfnamefont {P.}~\bibnamefont {Hurh}},
  \bibinfo {author} {\bibfnamefont {E.}~\bibnamefont {Dooryhee}}, \bibinfo
  {author} {\bibfnamefont {L.}~\bibnamefont {Snead}}, \bibinfo {author}
  {\bibfnamefont {D.}~\bibnamefont {Sprouster}}, \bibinfo {author}
  {\bibfnamefont {Z.}~\bibnamefont {Zhong}}, \bibinfo {author} {\bibfnamefont
  {H.}~\bibnamefont {Zhong}}, \bibinfo {author} {\bibfnamefont
  {S.}~\bibnamefont {Ghose}}, \bibinfo {author} {\bibfnamefont
  {Z.}~\bibnamefont {Kotsina}}, \bibinfo {author} {\bibfnamefont
  {K.}~\bibnamefont {Ammigan}}, \emph {et~al.},\ }\bibfield  {title} {\bibinfo
  {title} {120 {GeV} neutrino physics graphite target damage assessment using
  electron microscopy and high-energy x-ray diffraction},\ }\href@noop {}
  {\bibfield  {journal} {\bibinfo  {journal} {Physical Review Accelerators and
  Beams}\ }\textbf {\bibinfo {volume} {22}},\ \bibinfo {pages} {041001}
  (\bibinfo {year} {2019})}\BibitemShut {NoStop}%
\bibitem [{\citenamefont {Sola}\ \emph {et~al.}(2019)\citenamefont {Sola},
  \citenamefont {Calviani}, \citenamefont {Aberle}, \citenamefont {Ahdida},
  \citenamefont {Avigni}, \citenamefont {Battistin}, \citenamefont {Bianchi},
  \citenamefont {Burger}, \citenamefont {Descarrega}, \citenamefont {Espadanal}
  \emph {et~al.}}]{sola2019beam}%
  \BibitemOpen
  \bibfield  {author} {\bibinfo {author} {\bibfnamefont {E.~L.}\ \bibnamefont
  {Sola}}, \bibinfo {author} {\bibfnamefont {M.}~\bibnamefont {Calviani}},
  \bibinfo {author} {\bibfnamefont {O.}~\bibnamefont {Aberle}}, \bibinfo
  {author} {\bibfnamefont {C.}~\bibnamefont {Ahdida}}, \bibinfo {author}
  {\bibfnamefont {P.}~\bibnamefont {Avigni}}, \bibinfo {author} {\bibfnamefont
  {M.}~\bibnamefont {Battistin}}, \bibinfo {author} {\bibfnamefont
  {L.}~\bibnamefont {Bianchi}}, \bibinfo {author} {\bibfnamefont
  {S.}~\bibnamefont {Burger}}, \bibinfo {author} {\bibfnamefont {J.~B.}\
  \bibnamefont {Descarrega}}, \bibinfo {author} {\bibfnamefont {J.~C.}\
  \bibnamefont {Espadanal}}, \emph {et~al.},\ }\bibfield  {title} {\bibinfo
  {title} {Beam impact tests of a prototype target for the beam dump facility
  at {CERN}: Experimental setup and preliminary analysis of the online
  results},\ }\href@noop {} {\bibfield  {journal} {\bibinfo  {journal}
  {Physical Review Accelerators and Beams}\ }\textbf {\bibinfo {volume} {22}},\
  \bibinfo {pages} {123001} (\bibinfo {year} {2019})}\BibitemShut {NoStop}%
\bibitem [{\citenamefont {Chao}(1993)}]{chao1993physics}%
  \BibitemOpen
  \bibfield  {author} {\bibinfo {author} {\bibfnamefont {A.~W.}\ \bibnamefont
  {Chao}},\ }\href@noop {} {\emph {\bibinfo {title} {Physics of collective beam
  instabilities in high energy accelerators}}}\ (\bibinfo  {publisher}
  {Wiley},\ \bibinfo {year} {1993})\BibitemShut {NoStop}%
\bibitem [{\citenamefont {Diskansky}\ and\ \citenamefont
  {Pestrikov}(1997)}]{diskansky1997physics}%
  \BibitemOpen
  \bibfield  {author} {\bibinfo {author} {\bibfnamefont {N.}~\bibnamefont
  {Diskansky}}\ and\ \bibinfo {author} {\bibfnamefont {D.}~\bibnamefont
  {Pestrikov}},\ }\href@noop {} {\emph {\bibinfo {title} {The physics of
  intense beams and storage rings}}}\ (\bibinfo  {publisher} {Springer Science
  \& Business Media},\ \bibinfo {year} {1997})\BibitemShut {NoStop}%
\bibitem [{\citenamefont {Ng}(2006)}]{ng2006physics}%
  \BibitemOpen
  \bibfield  {author} {\bibinfo {author} {\bibfnamefont {K.-Y.}\ \bibnamefont
  {Ng}},\ }\href@noop {} {\emph {\bibinfo {title} {Physics of intensity
  dependent beam instabilities}}}\ (\bibinfo  {publisher} {World Scientific},\
  \bibinfo {year} {2006})\BibitemShut {NoStop}%
\bibitem [{\citenamefont {Reiser}(2008)}]{Reiser}%
  \BibitemOpen
  \bibfield  {author} {\bibinfo {author} {\bibfnamefont {M.}~\bibnamefont
  {Reiser}},\ }\href@noop {} {\emph {\bibinfo {title} {Theory and design of
  charged particle beams}}}\ (\bibinfo  {publisher} {John Wiley and Sons},\
  \bibinfo {year} {2008})\BibitemShut {NoStop}%
\bibitem [{\citenamefont {Hofmann}(2017)}]{hofmann2017space}%
  \BibitemOpen
  \bibfield  {author} {\bibinfo {author} {\bibfnamefont {I.}~\bibnamefont
  {Hofmann}},\ }\href@noop {} {\emph {\bibinfo {title} {Space charge physics
  for particle accelerators}}}\ (\bibinfo  {publisher} {Springer},\ \bibinfo
  {year} {2017})\BibitemShut {NoStop}%
\bibitem [{\citenamefont {Minty}\ and\ \citenamefont
  {Zimmermann}(2003)}]{minty2003measurement}%
  \BibitemOpen
  \bibfield  {author} {\bibinfo {author} {\bibfnamefont {M.}~\bibnamefont
  {Minty}}\ and\ \bibinfo {author} {\bibfnamefont {F.}~\bibnamefont
  {Zimmermann}},\ }\href@noop {} {\emph {\bibinfo {title} {Measurement and
  control of charged particle beams}}}\ (\bibinfo  {publisher} {Springer
  Nature},\ \bibinfo {year} {2003})\BibitemShut {NoStop}%
\bibitem [{\citenamefont {Strehl}(2006)}]{strehl2006beam}%
  \BibitemOpen
  \bibfield  {author} {\bibinfo {author} {\bibfnamefont {P.}~\bibnamefont
  {Strehl}},\ }\href@noop {} {\emph {\bibinfo {title} {Beam instrumentation and
  diagnostics}}},\ Vol.\ \bibinfo {volume} {120}\ (\bibinfo  {publisher}
  {Springer},\ \bibinfo {year} {2006})\BibitemShut {NoStop}%
\bibitem [{\citenamefont {Huang}\ \emph {et~al.}(2006)\citenamefont {Huang},
  \citenamefont {Lee}, \citenamefont {Ng},\ and\ \citenamefont
  {Su}}]{huang2006emittance}%
  \BibitemOpen
  \bibfield  {author} {\bibinfo {author} {\bibfnamefont {X.}~\bibnamefont
  {Huang}}, \bibinfo {author} {\bibfnamefont {S.}~\bibnamefont {Lee}}, \bibinfo
  {author} {\bibfnamefont {K.}~\bibnamefont {Ng}},\ and\ \bibinfo {author}
  {\bibfnamefont {Y.}~\bibnamefont {Su}},\ }\bibfield  {title} {\bibinfo
  {title} {Emittance measurement and modeling for the {Fermilab Booster}},\
  }\href@noop {} {\bibfield  {journal} {\bibinfo  {journal} {Physical Review
  Special Topics-Accelerators and Beams}\ }\textbf {\bibinfo {volume} {9}},\
  \bibinfo {pages} {014202} (\bibinfo {year} {2006})}\BibitemShut {NoStop}%
\bibitem [{\citenamefont {Wittenburg}(2013)}]{wittenburg2013instrum}%
  \BibitemOpen
  \bibfield  {author} {\bibinfo {author} {\bibfnamefont {K.}~\bibnamefont
  {Wittenburg}},\ }\bibfield  {title} {\bibinfo {title} {Specific
  instrumentation and diagnostics for high-intensity hadron beams},\
  }\href@noop {} {\bibfield  {journal} {\bibinfo  {journal} {arXiv preprint
  arXiv:1303.6767}\ } (\bibinfo {year} {2013})}\BibitemShut {NoStop}%
\bibitem [{\citenamefont {Eldred}(2019)}]{eldred2019physics}%
  \BibitemOpen
  \bibfield  {author} {\bibinfo {author} {\bibfnamefont {J.}~\bibnamefont
  {Eldred}},\ }\bibfield  {title} {\bibinfo {title} {Physics studies for high
  intensity proton beams at the {Fermilab Booster}},\ }in\ \href@noop {} {\emph
  {\bibinfo {booktitle} {Proc. of NAPAC'19 (Lansing, MI, USA, Sep.1-6,
  2019)}}}\ (\bibinfo {organization} {JACoW},\ \bibinfo {year}
  {2019})\BibitemShut {NoStop}%
\bibitem [{\citenamefont {Shiltsev}(2017)}]{ShiltsevMPLA}%
  \BibitemOpen
  \bibfield  {author} {\bibinfo {author} {\bibfnamefont {V.}~\bibnamefont
  {Shiltsev}},\ }\bibfield  {title} {\bibinfo {title} {Fermilab proton
  accelerator complex status and improvement plans},\ }\href@noop {} {\bibfield
   {journal} {\bibinfo  {journal} {Modern Physics Letters A}\ }\textbf
  {\bibinfo {volume} {32}},\ \bibinfo {pages} {1730012} (\bibinfo {year}
  {2017})}\BibitemShut {NoStop}%
\bibitem [{\citenamefont {Convery}\ \emph {et~al.}(2018)\citenamefont
  {Convery}, \citenamefont {Lindgren}, \citenamefont {Nagaitsev},\ and\
  \citenamefont {Shiltsev}}]{Convery}%
  \BibitemOpen
  \bibfield  {author} {\bibinfo {author} {\bibfnamefont {M.}~\bibnamefont
  {Convery}}, \bibinfo {author} {\bibfnamefont {M.}~\bibnamefont {Lindgren}},
  \bibinfo {author} {\bibfnamefont {S.}~\bibnamefont {Nagaitsev}},\ and\
  \bibinfo {author} {\bibfnamefont {V.}~\bibnamefont {Shiltsev}},\ }\href@noop
  {} {\emph {\bibinfo {title} {Fermilab accelerator complex: status and
  improvement plans}}},\ \bibinfo {type} {Tech. Rep.}\ \bibinfo {number}
  {FERMILAB-TM-2693}\ (\bibinfo  {institution} {Fermilab},\ \bibinfo {year}
  {2018})\BibitemShut {NoStop}%
\bibitem [{\citenamefont {Acciarri}\ \emph {et~al.}(2016)\citenamefont
  {Acciarri} \emph {et~al.}}]{dune2016lbnf}%
  \BibitemOpen
  \bibfield  {author} {\bibinfo {author} {\bibfnamefont {R.}~\bibnamefont
  {Acciarri}} \emph {et~al.},\ }\bibfield  {title} {\bibinfo {title}
  {Long-baseline neutrino facility {(LBNF)} and deep underground neutrino
  experiment {(DUNE)} conceptual design report volume 1: The {LBNF and DUNE}
  projects},\ }\href@noop {} {\bibfield  {journal} {\bibinfo  {journal} {arXiv
  preprint arXiv:1601.05471}\ } (\bibinfo {year} {2016})}\BibitemShut {NoStop}%
\bibitem [{\citenamefont {Lebedev}\ \emph {et~al.}(2017)\citenamefont {Lebedev}
  \emph {et~al.}}]{PIPII2017lebedev}%
  \BibitemOpen
  \bibfield  {author} {\bibinfo {author} {\bibfnamefont {V.}~\bibnamefont
  {Lebedev}} \emph {et~al.},\ }\bibfield  {title} {\bibinfo {title} {The
  {PIP-II} conceptual design report},\ }\href@noop {} {\bibfield  {journal}
  {\bibinfo  {journal} {Fermilab, Batavia, FERMILAB-TM-2649-AD-APC}\ }
  (\bibinfo {year} {2017})}\BibitemShut {NoStop}%
\bibitem [{\citenamefont {Eldred}\ \emph
  {et~al.}(2019{\natexlab{a}})\citenamefont {Eldred}, \citenamefont {Lebedev},\
  and\ \citenamefont {Valishev}}]{eldred2019pipiii}%
  \BibitemOpen
  \bibfield  {author} {\bibinfo {author} {\bibfnamefont {J.}~\bibnamefont
  {Eldred}}, \bibinfo {author} {\bibfnamefont {V.}~\bibnamefont {Lebedev}},\
  and\ \bibinfo {author} {\bibfnamefont {A.}~\bibnamefont {Valishev}},\
  }\bibfield  {title} {\bibinfo {title} {Rapid-cycling synchrotron for
  multi-megawatt proton facility at fermilab},\ }\href@noop {} {\bibfield
  {journal} {\bibinfo  {journal} {Journal of Instrumentation}\ }\textbf
  {\bibinfo {volume} {14}}\bibinfo  {number} { (07)},\ \bibinfo {pages}
  {P07021}}\BibitemShut {NoStop}%
\bibitem [{\citenamefont {Eldred}\ \emph {et~al.}(2020)\citenamefont {Eldred}
  \emph {et~al.}}]{eldred2020snowmass}%
  \BibitemOpen
\bibfield  {number} {  }\bibfield  {author} {\bibinfo {author} {\bibfnamefont
  {J.}~\bibnamefont {Eldred}} \emph {et~al.},\ }\bibfield  {title} {\bibinfo
  {title} {Versatile multi-mw proton facility with synchrotron upgrade of
  fermilab proton complex},\ }in\ \href@noop {} {\emph {\bibinfo {booktitle}
  {Snowmass 2021 Letters of Interest}}}\ (\bibinfo {organization} {APS DPF},\
  \bibinfo {year} {2020})\BibitemShut {NoStop}%
\bibitem [{\citenamefont {Hubbard}\ \emph {et~al.}(1973)\citenamefont {Hubbard}
  \emph {et~al.}}]{Booster}%
  \BibitemOpen
  \bibfield  {author} {\bibinfo {author} {\bibfnamefont {E.}~\bibnamefont
  {Hubbard}} \emph {et~al.},\ }\href@noop {} {\emph {\bibinfo {title} {Booster
  synchrotron}}},\ \bibinfo {type} {Tech. Rep.}\ \bibinfo {number}
  {FERMILAB-TM-405}\ (\bibinfo  {institution} {Fermilab},\ \bibinfo {year}
  {1973})\BibitemShut {NoStop}%
\bibitem [{Boo()}]{BoosterBook}%
  \BibitemOpen
  \href@noop {} {\bibinfo {title} {Booster rookie book}},\ \bibinfo
  {howpublished}
  {\url{https://operations.fnal.gov/rookie_books/Booster_V4.1.pdf}},\ \bibinfo
  {note} {accessed: March 3, 2020}\BibitemShut {NoStop}%
\bibitem [{\citenamefont {Eldred}(2015)}]{eldred2015phd}%
  \BibitemOpen
  \bibfield  {author} {\bibinfo {author} {\bibfnamefont {J.}~\bibnamefont
  {Eldred}},\ }\emph {\bibinfo {title} {Slip-stacking Dynamics for High-Power
  Proton Beams at Fermilab}},\ \href@noop {} {Ph.D. thesis},\ \bibinfo
  {school} {U. Indiana, Tech.Rep. FERMILAB-THESIS-2015-31} (\bibinfo {year}
  {2015})\BibitemShut {NoStop}%
\bibitem [{\citenamefont {Seiya}\ \emph {et~al.}(2015)\citenamefont {Seiya},
  \citenamefont {Bhat}, \citenamefont {Johnson}, \citenamefont {Kapin},
  \citenamefont {Pellico}, \citenamefont {Tan},\ and\ \citenamefont
  {Tesarek}}]{seiya2015beam}%
  \BibitemOpen
  \bibfield  {author} {\bibinfo {author} {\bibfnamefont {K.}~\bibnamefont
  {Seiya}}, \bibinfo {author} {\bibfnamefont {C.}~\bibnamefont {Bhat}},
  \bibinfo {author} {\bibfnamefont {D.}~\bibnamefont {Johnson}}, \bibinfo
  {author} {\bibfnamefont {V.}~\bibnamefont {Kapin}}, \bibinfo {author}
  {\bibfnamefont {W.}~\bibnamefont {Pellico}}, \bibinfo {author} {\bibfnamefont
  {C.-Y.}\ \bibnamefont {Tan}},\ and\ \bibinfo {author} {\bibfnamefont
  {R.}~\bibnamefont {Tesarek}},\ }\bibfield  {title} {\bibinfo {title} {Beam
  studies for the {Proton Improvement Plan (PIP)}--reducing beam loss at the
  {Fermilab Booster}},\ }\href@noop {} {\bibfield  {journal} {\bibinfo
  {journal} {arXiv preprint arXiv:1511.01467}\ } (\bibinfo {year}
  {2015})}\BibitemShut {NoStop}%
\bibitem [{\citenamefont {Budker}\ and\ \citenamefont
  {Dimov}(1964)}]{budkerdimov1963stripping}%
  \BibitemOpen
  \bibfield  {author} {\bibinfo {author} {\bibfnamefont {G.}~\bibnamefont
  {Budker}}\ and\ \bibinfo {author} {\bibfnamefont {G.}~\bibnamefont {Dimov}},\
  }\bibfield  {title} {\bibinfo {title} {On the charge exchange injection of
  protons into ring accelerators},\ }\href@noop {} {\bibfield  {journal}
  {\bibinfo  {journal} {Transactions of the International Conference on
  Accelerators (Dubna, 1963)}\ ,\ \bibinfo {pages} {993}} (\bibinfo {year}
  {1964})}\BibitemShut {NoStop}%
\bibitem [{\citenamefont {Hojvat}\ \emph {et~al.}(1979)\citenamefont {Hojvat},
  \citenamefont {Ankenbrandt}, \citenamefont {Brown}, \citenamefont {Cosgrove},
  \citenamefont {Garvey}, \citenamefont {Johnson}, \citenamefont {Joy},
  \citenamefont {Lackey}, \citenamefont {Meisner}, \citenamefont {Schmitz}
  \emph {et~al.}}]{hojvat1979multiturn}%
  \BibitemOpen
  \bibfield  {author} {\bibinfo {author} {\bibfnamefont {C.}~\bibnamefont
  {Hojvat}}, \bibinfo {author} {\bibfnamefont {C.}~\bibnamefont {Ankenbrandt}},
  \bibinfo {author} {\bibfnamefont {B.}~\bibnamefont {Brown}}, \bibinfo
  {author} {\bibfnamefont {D.}~\bibnamefont {Cosgrove}}, \bibinfo {author}
  {\bibfnamefont {J.}~\bibnamefont {Garvey}}, \bibinfo {author} {\bibfnamefont
  {R.}~\bibnamefont {Johnson}}, \bibinfo {author} {\bibfnamefont
  {M.}~\bibnamefont {Joy}}, \bibinfo {author} {\bibfnamefont {J.}~\bibnamefont
  {Lackey}}, \bibinfo {author} {\bibfnamefont {K.}~\bibnamefont {Meisner}},
  \bibinfo {author} {\bibfnamefont {T.}~\bibnamefont {Schmitz}}, \emph
  {et~al.},\ }\bibfield  {title} {\bibinfo {title} {The multiturn charge
  exchange injection system for the {Fermilab} booster accelerator},\
  }\href@noop {} {\bibfield  {journal} {\bibinfo  {journal} {IEEE Transactions
  on Nuclear Science}\ }\textbf {\bibinfo {volume} {26}},\ \bibinfo {pages}
  {3149} (\bibinfo {year} {1979})}\BibitemShut {NoStop}%
\bibitem [{\citenamefont {Bhat}(2015)}]{bhat2015newinj}%
  \BibitemOpen
  \bibfield  {author} {\bibinfo {author} {\bibfnamefont {C.}~\bibnamefont
  {Bhat}},\ }\bibfield  {title} {\bibinfo {title} {A new beam injection scheme
  for the {Fermilab Booster}},\ }\href@noop {} {\bibfield  {journal} {\bibinfo
  {journal} {arXiv preprint arXiv:1504.07174}\ } (\bibinfo {year}
  {2015})}\BibitemShut {NoStop}%
\bibitem [{\citenamefont {Bhat}(2017)}]{bhat2017injstudies}%
  \BibitemOpen
  \bibfield  {author} {\bibinfo {author} {\bibfnamefont {C.}~\bibnamefont
  {Bhat}},\ }\bibfield  {title} {\bibinfo {title} {{R\&D} on beam injection and
  bunching schemes in the {Fermilab Booster}},\ }\href@noop {} {\bibfield
  {journal} {\bibinfo  {journal} {arXiv preprint arXiv:1704.08157}\ } (\bibinfo
  {year} {2017})}\BibitemShut {NoStop}%
\bibitem [{\citenamefont {Macridin}\ \emph {et~al.}(2011)\citenamefont
  {Macridin}, \citenamefont {Spentzouris}, \citenamefont {Amundson},
  \citenamefont {Spentzouris},\ and\ \citenamefont
  {McCarron}}]{macridin2011coupling}%
  \BibitemOpen
  \bibfield  {author} {\bibinfo {author} {\bibfnamefont {A.}~\bibnamefont
  {Macridin}}, \bibinfo {author} {\bibfnamefont {P.}~\bibnamefont
  {Spentzouris}}, \bibinfo {author} {\bibfnamefont {J.}~\bibnamefont
  {Amundson}}, \bibinfo {author} {\bibfnamefont {L.}~\bibnamefont
  {Spentzouris}},\ and\ \bibinfo {author} {\bibfnamefont {D.}~\bibnamefont
  {McCarron}},\ }\bibfield  {title} {\bibinfo {title} {Coupling impedance and
  wake functions for laminated structures with an application to the fermilab
  booster},\ }\href@noop {} {\bibfield  {journal} {\bibinfo  {journal}
  {Physical Review Special Topics-Accelerators and Beams}\ }\textbf {\bibinfo
  {volume} {14}},\ \bibinfo {pages} {061003} (\bibinfo {year}
  {2011})}\BibitemShut {NoStop}%
\bibitem [{\citenamefont {Macridin}\ \emph {et~al.}(2013)\citenamefont
  {Macridin}, \citenamefont {Amundson}, \citenamefont {Spentzouris},
  \citenamefont {Lebedev},\ and\ \citenamefont
  {Zolkin}}]{macridin2013transverse}%
  \BibitemOpen
  \bibfield  {author} {\bibinfo {author} {\bibfnamefont {A.}~\bibnamefont
  {Macridin}}, \bibinfo {author} {\bibfnamefont {J.}~\bibnamefont {Amundson}},
  \bibinfo {author} {\bibfnamefont {P.}~\bibnamefont {Spentzouris}}, \bibinfo
  {author} {\bibfnamefont {V.}~\bibnamefont {Lebedev}},\ and\ \bibinfo {author}
  {\bibfnamefont {T.}~\bibnamefont {Zolkin}},\ }\href@noop {} {\emph {\bibinfo
  {title} {Transverse Impedance and Transverse Instabilities in Fermilab
  Booster}}},\ \bibinfo {type} {Tech. Rep.}\ \bibinfo {number}
  {FERMILAB-CONF-13-431}\ (\bibinfo  {institution} {Fermilab},\ \bibinfo {year}
  {2013})\BibitemShut {NoStop}%
\bibitem [{\citenamefont {Valishev}\ \emph {et~al.}(2016)\citenamefont
  {Valishev}, \citenamefont {Alexahin},\ and\ \citenamefont
  {Lebedev}}]{valishev2016suppression}%
  \BibitemOpen
  \bibfield  {author} {\bibinfo {author} {\bibfnamefont {A.}~\bibnamefont
  {Valishev}}, \bibinfo {author} {\bibfnamefont {Y.}~\bibnamefont {Alexahin}},\
  and\ \bibinfo {author} {\bibfnamefont {V.}~\bibnamefont {Lebedev}},\
  }\bibfield  {title} {\bibinfo {title} {Suppression of half-integer resonance
  in {FNAL Booster} and space charge losses at injection},\ }in\ \href@noop {}
  {\emph {\bibinfo {booktitle} {Proc. HB2016 (Malmö, Sweden)}}}\ (\bibinfo
  {organization} {JACoW},\ \bibinfo {year} {2016})\ pp.\ \bibinfo {pages}
  {164--168}\BibitemShut {NoStop}%
\bibitem [{\citenamefont {Yang}\ \emph {et~al.}(2005)\citenamefont {Yang},
  \citenamefont {Ankenbrandt}, \citenamefont {MacLachlan},\ and\ \citenamefont
  {Lebedev}}]{yang2005trans}%
  \BibitemOpen
  \bibfield  {author} {\bibinfo {author} {\bibfnamefont {X.}~\bibnamefont
  {Yang}}, \bibinfo {author} {\bibfnamefont {C.~M.}\ \bibnamefont
  {Ankenbrandt}}, \bibinfo {author} {\bibfnamefont {J.}~\bibnamefont
  {MacLachlan}},\ and\ \bibinfo {author} {\bibfnamefont {V.~A.}\ \bibnamefont
  {Lebedev}},\ }\href@noop {} {\emph {\bibinfo {title} {A proposed transition
  scheme for the longitudinal emittance control in the Fermilab Booster}}},\
  \bibinfo {type} {Tech. Rep.}\ \bibinfo {number} {FERMILAB-FN-0772}\ (\bibinfo
   {institution} {Fermilab},\ \bibinfo {year} {2005})\BibitemShut {NoStop}%
\bibitem [{\citenamefont {Lebedev}\ \emph {et~al.}(2016)\citenamefont
  {Lebedev}, \citenamefont {Ostiguy},\ and\ \citenamefont
  {Bhat}}]{lebedev2016trans}%
  \BibitemOpen
  \bibfield  {author} {\bibinfo {author} {\bibfnamefont {V.}~\bibnamefont
  {Lebedev}}, \bibinfo {author} {\bibfnamefont {J.-F.}\ \bibnamefont
  {Ostiguy}},\ and\ \bibinfo {author} {\bibfnamefont {C.}~\bibnamefont
  {Bhat}},\ }\bibfield  {title} {\bibinfo {title} {Beam acceleration and
  transition crossing in the fermilab booster},\ }in\ \href@noop {} {\emph
  {\bibinfo {booktitle} {Proc. HB2016 (Malmö, Sweden)}}}\ (\bibinfo
  {organization} {JACoW},\ \bibinfo {year} {2016})\ pp.\ \bibinfo {pages}
  {160--163}\BibitemShut {NoStop}%
\bibitem [{\citenamefont {Ostiguy}\ \emph {et~al.}(2016)\citenamefont
  {Ostiguy}, \citenamefont {Bhat},\ and\ \citenamefont
  {Lebedev}}]{ostiguy2016trans}%
  \BibitemOpen
  \bibfield  {author} {\bibinfo {author} {\bibfnamefont {J.-F.}\ \bibnamefont
  {Ostiguy}}, \bibinfo {author} {\bibfnamefont {C.}~\bibnamefont {Bhat}},\ and\
  \bibinfo {author} {\bibfnamefont {V.}~\bibnamefont {Lebedev}},\ }\bibfield
  {title} {\bibinfo {title} {Modeling longitudinal dynamics in the fermilab
  booster synchrotron},\ }in\ \href@noop {} {\emph {\bibinfo {booktitle} {Proc.
  Proc. IPAC2016 (Busan, Korea)}}}\ (\bibinfo {organization} {JACoW},\ \bibinfo
  {year} {2016})\ pp.\ \bibinfo {pages} {873--876}\BibitemShut {NoStop}%
\bibitem [{\citenamefont {Yang}\ \emph {et~al.}(2007)\citenamefont {Yang},
  \citenamefont {Drozhdin},\ and\ \citenamefont {Pellico}}]{yang2007momentum}%
  \BibitemOpen
  \bibfield  {author} {\bibinfo {author} {\bibfnamefont {X.}~\bibnamefont
  {Yang}}, \bibinfo {author} {\bibfnamefont {A.}~\bibnamefont {Drozhdin}},\
  and\ \bibinfo {author} {\bibfnamefont {W.}~\bibnamefont {Pellico}},\
  }\bibfield  {title} {\bibinfo {title} {Momentum spread reduction at beam
  extraction from the fermilab booster at slipstacking injection to the main
  injector},\ }in\ \href@noop {} {\emph {\bibinfo {booktitle} {2007 IEEE
  Particle Accelerator Conference (PAC)}}}\ (\bibinfo {organization} {IEEE},\
  \bibinfo {year} {2007})\ pp.\ \bibinfo {pages} {1733--1735}\BibitemShut
  {NoStop}%
\bibitem [{acn()}]{acnet2019}%
  \BibitemOpen
  \href@noop {} {\bibinfo {title} {{B.Hendricks, ACNET: The Undiscovered
  Control System}}},\ \bibinfo {howpublished} {Beams-doc-7037-v2},\ \bibinfo
  {note} {(Fermilab internal note)}\BibitemShut {NoStop}%
\bibitem [{\citenamefont {Shiltsev}\ \emph
  {et~al.}(2020{\natexlab{a}})\citenamefont {Shiltsev}, \citenamefont {Eldred},
  \citenamefont {Lebedev},\ and\ \citenamefont {Seiya}}]{Shiltsev2020a}%
  \BibitemOpen
  \bibfield  {author} {\bibinfo {author} {\bibfnamefont {V.}~\bibnamefont
  {Shiltsev}}, \bibinfo {author} {\bibfnamefont {J.}~\bibnamefont {Eldred}},
  \bibinfo {author} {\bibfnamefont {V.}~\bibnamefont {Lebedev}},\ and\ \bibinfo
  {author} {\bibfnamefont {K.}~\bibnamefont {Seiya}},\ }\href@noop {} {\emph
  {\bibinfo {title} {Studies of Beam Intensity Effects in the {Fermilab
  Booster} Synchrotron. {Part I}: Introduction; Tune and Chromaticity Scans of
  Beam Losses}}},\ \bibinfo {type} {Tech. Rep.}\ \bibinfo {number}
  {FERMILAB-TM-2740}\ (\bibinfo  {institution} {Fermilab},\ \bibinfo {year}
  {2020})\BibitemShut {NoStop}%
\bibitem [{\citenamefont {Johnson}\ \emph {et~al.}(2018)\citenamefont
  {Johnson}, \citenamefont {Johnson}, \citenamefont {Bhat}, \citenamefont
  {Chaurize}, \citenamefont {Duel}, \citenamefont {Karns}, \citenamefont
  {Pellico}, \citenamefont {Schupbach}, \citenamefont {Seiya},\ and\
  \citenamefont {Slimmer}}]{johnson2018mebt}%
  \BibitemOpen
  \bibfield  {author} {\bibinfo {author} {\bibfnamefont {D.}~\bibnamefont
  {Johnson}}, \bibinfo {author} {\bibfnamefont {T.}~\bibnamefont {Johnson}},
  \bibinfo {author} {\bibfnamefont {C.}~\bibnamefont {Bhat}}, \bibinfo {author}
  {\bibfnamefont {S.}~\bibnamefont {Chaurize}}, \bibinfo {author}
  {\bibfnamefont {K.}~\bibnamefont {Duel}}, \bibinfo {author} {\bibfnamefont
  {P.}~\bibnamefont {Karns}}, \bibinfo {author} {\bibfnamefont
  {W.}~\bibnamefont {Pellico}}, \bibinfo {author} {\bibfnamefont
  {B.}~\bibnamefont {Schupbach}}, \bibinfo {author} {\bibfnamefont
  {K.}~\bibnamefont {Seiya}},\ and\ \bibinfo {author} {\bibfnamefont
  {D.}~\bibnamefont {Slimmer}},\ }\bibfield  {title} {\bibinfo {title} {Mebt
  laser notcher (chopper) for {Booster} loss reduction},\ }in\ \href@noop {}
  {\emph {\bibinfo {booktitle} {ICFA ABDW on High-Intensity and High-Brightness
  Hadron Beams (Daejeon, Korea, 17-22 June 2018)}}}\ (\bibinfo {organization}
  {JACOW},\ \bibinfo {year} {2018})\ pp.\ \bibinfo {pages}
  {416--421}\BibitemShut {NoStop}%
\bibitem [{\citenamefont {Burov}\ and\ \citenamefont
  {Lebedev}(2012)}]{burov2012laminated}%
  \BibitemOpen
  \bibfield  {author} {\bibinfo {author} {\bibfnamefont {A.}~\bibnamefont
  {Burov}}\ and\ \bibinfo {author} {\bibfnamefont {V.}~\bibnamefont
  {Lebedev}},\ }\bibfield  {title} {\bibinfo {title} {Impedances of laminated
  vacuum chambers},\ }\href@noop {} {\bibfield  {journal} {\bibinfo  {journal}
  {arXiv preprint arXiv:1209.2996}\ } (\bibinfo {year} {2012})}\BibitemShut
  {NoStop}%
\bibitem [{\citenamefont {Shiltsev}\ \emph {et~al.}(2005)\citenamefont
  {Shiltsev}, \citenamefont {Alexahin}, \citenamefont {Lebedev}, \citenamefont
  {Lebrun}, \citenamefont {Moore}, \citenamefont {Sen}, \citenamefont
  {Tollestrup}, \citenamefont {Valishev},\ and\ \citenamefont
  {Zhang}}]{shiltsev2005beambeam}%
  \BibitemOpen
  \bibfield  {author} {\bibinfo {author} {\bibfnamefont {V.}~\bibnamefont
  {Shiltsev}}, \bibinfo {author} {\bibfnamefont {Y.}~\bibnamefont {Alexahin}},
  \bibinfo {author} {\bibfnamefont {V.}~\bibnamefont {Lebedev}}, \bibinfo
  {author} {\bibfnamefont {P.}~\bibnamefont {Lebrun}}, \bibinfo {author}
  {\bibfnamefont {R.}~\bibnamefont {Moore}}, \bibinfo {author} {\bibfnamefont
  {T.}~\bibnamefont {Sen}}, \bibinfo {author} {\bibfnamefont {A.}~\bibnamefont
  {Tollestrup}}, \bibinfo {author} {\bibfnamefont {A.}~\bibnamefont
  {Valishev}},\ and\ \bibinfo {author} {\bibfnamefont {X.}~\bibnamefont
  {Zhang}},\ }\bibfield  {title} {\bibinfo {title} {Beam-beam effects in the
  tevatron},\ }\href@noop {} {\bibfield  {journal} {\bibinfo  {journal}
  {Physical Review Special Topics-Accelerators and Beams}\ }\textbf {\bibinfo
  {volume} {8}},\ \bibinfo {pages} {101001} (\bibinfo {year}
  {2005})}\BibitemShut {NoStop}%
\bibitem [{\citenamefont {Zagel}\ \emph {et~al.}(2010)\citenamefont {Zagel},
  \citenamefont {Jansson}, \citenamefont {Meyer}, \citenamefont {Morris},
  \citenamefont {Slimmer}, \citenamefont {Sullivan},\ and\ \citenamefont
  {Yang}}]{zagel2010ipms}%
  \BibitemOpen
  \bibfield  {author} {\bibinfo {author} {\bibfnamefont {J.}~\bibnamefont
  {Zagel}}, \bibinfo {author} {\bibfnamefont {A.}~\bibnamefont {Jansson}},
  \bibinfo {author} {\bibfnamefont {T.}~\bibnamefont {Meyer}}, \bibinfo
  {author} {\bibfnamefont {D.~K.}\ \bibnamefont {Morris}}, \bibinfo {author}
  {\bibfnamefont {D.}~\bibnamefont {Slimmer}}, \bibinfo {author} {\bibfnamefont
  {T.}~\bibnamefont {Sullivan}},\ and\ \bibinfo {author} {\bibfnamefont
  {M.}~\bibnamefont {Yang}},\ }\bibfield  {title} {\bibinfo {title}
  {Operational use of ionization profile monitors at {Fermilab}},\ }in\
  \href@noop {} {\emph {\bibinfo {booktitle} {BIW2010, (Santa Fe, New Mexico,
  US)}}}\ (\bibinfo {organization} {JACOW},\ \bibinfo {year}
  {2010})\BibitemShut {NoStop}%
\bibitem [{\citenamefont {Shiltsev}(2020{\natexlab{c}})}]{shiltsev2020ipm}%
  \BibitemOpen
  \bibfield  {author} {\bibinfo {author} {\bibfnamefont {V.}~\bibnamefont
  {Shiltsev}},\ }\bibfield  {title} {\bibinfo {title} {Space-charge effects in
  ionization beam profile monitors},\ }\href@noop {} {\bibfield  {journal}
  {\bibinfo  {journal} {Nuclear Instruments and Methods in Physics Research
  Section A: Accelerators, Spectrometers, Detectors and Associated Equipment}\
  ,\ \bibinfo {pages} {164744}} (\bibinfo {year} {2020}{\natexlab{c}})},\
  \bibinfo {note} {also in arXiv:2003.09072}\BibitemShut {NoStop}%
\bibitem [{\citenamefont {Shiltsev}\ \emph
  {et~al.}(2020{\natexlab{b}})\citenamefont {Shiltsev}, \citenamefont {Eldred},
  \citenamefont {Lebedev},\ and\ \citenamefont {Seiya}}]{Shiltsev2020}%
  \BibitemOpen
  \bibfield  {author} {\bibinfo {author} {\bibfnamefont {V.}~\bibnamefont
  {Shiltsev}}, \bibinfo {author} {\bibfnamefont {J.}~\bibnamefont {Eldred}},
  \bibinfo {author} {\bibfnamefont {V.}~\bibnamefont {Lebedev}},\ and\ \bibinfo
  {author} {\bibfnamefont {K.}~\bibnamefont {Seiya}},\ }\href@noop {} {\emph
  {\bibinfo {title} {Studies of Beam Intensity Effects in the {Fermilab
  Booster} Synchrotron. {Part II}: Beam Emittance Evolution}}},\ \bibinfo
  {type} {Tech. Rep.}\ \bibinfo {number} {FERMILAB-TM-2741}\ (\bibinfo
  {institution} {Fermilab},\ \bibinfo {year} {2020})\BibitemShut {NoStop}%
\bibitem [{\citenamefont {Bhat}\ \emph {et~al.}(2015)\citenamefont {Bhat},
  \citenamefont {Chase}, \citenamefont {Chaurize}, \citenamefont {Garcia},
  \citenamefont {Seiya}, \citenamefont {Pellico}, \citenamefont {Sullivan},\
  and\ \citenamefont {Triplett}}]{bhat2015injemm}%
  \BibitemOpen
  \bibfield  {author} {\bibinfo {author} {\bibfnamefont {C.}~\bibnamefont
  {Bhat}}, \bibinfo {author} {\bibfnamefont {B.}~\bibnamefont {Chase}},
  \bibinfo {author} {\bibfnamefont {S.}~\bibnamefont {Chaurize}}, \bibinfo
  {author} {\bibfnamefont {F.}~\bibnamefont {Garcia}}, \bibinfo {author}
  {\bibfnamefont {K.}~\bibnamefont {Seiya}}, \bibinfo {author} {\bibfnamefont
  {W.}~\bibnamefont {Pellico}}, \bibinfo {author} {\bibfnamefont
  {T.}~\bibnamefont {Sullivan}},\ and\ \bibinfo {author} {\bibfnamefont
  {A.~K.}\ \bibnamefont {Triplett}},\ }\bibfield  {title} {\bibinfo {title}
  {Energy spread of the proton beam in the {Fermilab Booster} at its injection
  energy},\ }\href@noop {} {\bibfield  {journal} {\bibinfo  {journal} {arXiv
  preprint arXiv:1504.07195}\ } (\bibinfo {year} {2015})}\BibitemShut {NoStop}%
\bibitem [{\citenamefont {Eldred}\ \emph
  {et~al.}(2019{\natexlab{b}})\citenamefont {Eldred}, \citenamefont {Bhat},
  \citenamefont {Chaurize}, \citenamefont {Lebedev}, \citenamefont {Nagaitsev},
  \citenamefont {Seiya}, \citenamefont {Tan},\ and\ \citenamefont
  {Tesarek}}]{eldred2019foil}%
  \BibitemOpen
  \bibfield  {author} {\bibinfo {author} {\bibfnamefont {J.}~\bibnamefont
  {Eldred}}, \bibinfo {author} {\bibfnamefont {C.}~\bibnamefont {Bhat}},
  \bibinfo {author} {\bibfnamefont {S.}~\bibnamefont {Chaurize}}, \bibinfo
  {author} {\bibfnamefont {V.}~\bibnamefont {Lebedev}}, \bibinfo {author}
  {\bibfnamefont {S.}~\bibnamefont {Nagaitsev}}, \bibinfo {author}
  {\bibfnamefont {K.}~\bibnamefont {Seiya}}, \bibinfo {author} {\bibfnamefont
  {C.}~\bibnamefont {Tan}},\ and\ \bibinfo {author} {\bibfnamefont
  {R.}~\bibnamefont {Tesarek}},\ }\bibfield  {title} {\bibinfo {title} {Foil
  scattering model for {Fermilab Booster}},\ }\href@noop {} {\bibfield
  {journal} {\bibinfo  {journal} {arXiv preprint arXiv:1912.02896}\ } (\bibinfo
  {year} {2019}{\natexlab{b}})}\BibitemShut {NoStop}%
\bibitem [{eld()}]{eldred2020ecloudstudies}%
  \BibitemOpen
  \href@noop {} {\bibinfo {title} {{J.Eldred, Preliminary double-notch Booster
  ecloud study}}},\ \bibinfo {howpublished} {Beams-doc-8910},\ \bibinfo {note}
  {(Fermilab internal note)}\BibitemShut {NoStop}%
\bibitem [{wei()}]{weiren1999icfa}%
  \BibitemOpen
  \href@noop {} {\bibinfo {title} {{W.Chou, Ch.4.14 in \textit{ICFA Beam
  Dynamics Newsletter No.20} (1999)}}},\ \bibinfo {howpublished}
  {\url{https://icfa-usa.jlab.org/archive/newsletter/icfa_bd_nl_20.pdf}},\
  \bibinfo {note} {accessed: November 16, 2020}\BibitemShut {NoStop}%
\bibitem [{\citenamefont {Tang}(2013)}]{tang2013rcs}%
  \BibitemOpen
  \bibfield  {author} {\bibinfo {author} {\bibfnamefont {J.}~\bibnamefont
  {Tang}},\ }\bibfield  {title} {\bibinfo {title} {Rapid cycling synchrotrons
  and accumulator rings for high-intensity hadron beams},\ }\href@noop {}
  {\bibfield  {journal} {\bibinfo  {journal} {Reviews of Accelerator Science
  and Technology}\ }\textbf {\bibinfo {volume} {6}},\ \bibinfo {pages} {143}
  (\bibinfo {year} {2013})}\BibitemShut {NoStop}%
\bibitem [{\citenamefont {Moore}\ \emph {et~al.}(1981)\citenamefont {Moore},
  \citenamefont {Curtis}, \citenamefont {Lackey}, \citenamefont {Owen},
  \citenamefont {Ankenbrandt}, \citenamefont {Gerig},\ and\ \citenamefont
  {Pruss}}]{moore1981dependence}%
  \BibitemOpen
  \bibfield  {author} {\bibinfo {author} {\bibfnamefont {C.}~\bibnamefont
  {Moore}}, \bibinfo {author} {\bibfnamefont {C.}~\bibnamefont {Curtis}},
  \bibinfo {author} {\bibfnamefont {J.}~\bibnamefont {Lackey}}, \bibinfo
  {author} {\bibfnamefont {C.}~\bibnamefont {Owen}}, \bibinfo {author}
  {\bibfnamefont {C.}~\bibnamefont {Ankenbrandt}}, \bibinfo {author}
  {\bibfnamefont {R.}~\bibnamefont {Gerig}},\ and\ \bibinfo {author}
  {\bibfnamefont {S.}~\bibnamefont {Pruss}},\ }\bibfield  {title} {\bibinfo
  {title} {Dependence of the emittances of the fermilab injectors on
  intensity},\ }\href@noop {} {\bibfield  {journal} {\bibinfo  {journal} {IEEE
  Transactions on Nuclear Science}\ }\textbf {\bibinfo {volume} {28}},\
  \bibinfo {pages} {3000} (\bibinfo {year} {1981})}\BibitemShut {NoStop}%
\bibitem [{\citenamefont {Popovic}\ and\ \citenamefont
  {Ankenbrandt}(1998)}]{popovic1998performance}%
  \BibitemOpen
  \bibfield  {author} {\bibinfo {author} {\bibfnamefont {M.}~\bibnamefont
  {Popovic}}\ and\ \bibinfo {author} {\bibfnamefont {C.}~\bibnamefont
  {Ankenbrandt}},\ }\bibfield  {title} {\bibinfo {title} {Performance and
  measurements of the fermilab booster},\ }in\ \href@noop {} {\emph {\bibinfo
  {booktitle} {AIP Conference Proceedings}}},\ Vol.\ \bibinfo {volume} {448}\
  (\bibinfo {organization} {American Institute of Physics},\ \bibinfo {year}
  {1998})\ pp.\ \bibinfo {pages} {128--134}\BibitemShut {NoStop}%
\bibitem [{\citenamefont {Chou}\ \emph {et~al.}(2003)\citenamefont {Chou},
  \citenamefont {Drozhdin}, \citenamefont {Lucas},\ and\ \citenamefont
  {Ostiguy}}]{chou2003fermilab}%
  \BibitemOpen
  \bibfield  {author} {\bibinfo {author} {\bibfnamefont {W.}~\bibnamefont
  {Chou}}, \bibinfo {author} {\bibfnamefont {A.}~\bibnamefont {Drozhdin}},
  \bibinfo {author} {\bibfnamefont {P.}~\bibnamefont {Lucas}},\ and\ \bibinfo
  {author} {\bibfnamefont {F.}~\bibnamefont {Ostiguy}},\ }\bibfield  {title}
  {\bibinfo {title} {Fermilab booster modeling and space charge study},\ }in\
  \href@noop {} {\emph {\bibinfo {booktitle} {Proceedings of the 2003 Particle
  Accelerator Conference}}},\ Vol.~\bibinfo {volume} {5}\ (\bibinfo
  {organization} {IEEE},\ \bibinfo {year} {2003})\ pp.\ \bibinfo {pages}
  {2925--2927}\BibitemShut {NoStop}%
\bibitem [{\citenamefont {Ankenbrandt}\ and\ \citenamefont
  {Holmes}(1987)}]{ankenbrandt1987limits}%
  \BibitemOpen
  \bibfield  {author} {\bibinfo {author} {\bibfnamefont {C.}~\bibnamefont
  {Ankenbrandt}}\ and\ \bibinfo {author} {\bibfnamefont {S.}~\bibnamefont
  {Holmes}},\ }\href@noop {} {\emph {\bibinfo {title} {Limits on the transverse
  phase space density in the Fermilab Booster}}},\ \bibinfo {type} {Tech.
  Rep.}\ (\bibinfo  {institution} {Fermi National Accelerator Lab.},\ \bibinfo
  {year} {1987})\BibitemShut {NoStop}%
\bibitem [{\citenamefont {Holmes}(1997)}]{holmes1997maininj}%
  \BibitemOpen
  \bibfield  {author} {\bibinfo {author} {\bibfnamefont {S.~D.}\ \bibnamefont
  {Holmes}},\ }\bibfield  {title} {\bibinfo {title} {Design criteria and
  performance goals for the fermilab main injector},\ }\href@noop {} {\bibfield
   {journal} {\bibinfo  {journal} {Part. Accel.}\ }\textbf {\bibinfo {volume}
  {58}},\ \bibinfo {pages} {39} (\bibinfo {year} {1997})}\BibitemShut {NoStop}%
\bibitem [{\citenamefont {Graves}\ \emph {et~al.}(1995)\citenamefont {Graves},
  \citenamefont {Bharadwaj},\ and\ \citenamefont {McGinnis}}]{graves1995ipm}%
  \BibitemOpen
  \bibfield  {author} {\bibinfo {author} {\bibfnamefont {W.~S.}\ \bibnamefont
  {Graves}}, \bibinfo {author} {\bibfnamefont {V.}~\bibnamefont {Bharadwaj}},\
  and\ \bibinfo {author} {\bibfnamefont {D.}~\bibnamefont {McGinnis}},\
  }\bibfield  {title} {\bibinfo {title} {A nondestructive fast beam profile
  monitor},\ }\href@noop {} {\bibfield  {journal} {\bibinfo  {journal} {Nuclear
  Instruments and Methods in Physics Research Section A: Accelerators,
  Spectrometers, Detectors and Associated Equipment}\ }\textbf {\bibinfo
  {volume} {364}},\ \bibinfo {pages} {13} (\bibinfo {year} {1995})}\BibitemShut
  {NoStop}%
\bibitem [{\citenamefont {Amundson}\ \emph {et~al.}(2003)\citenamefont
  {Amundson}, \citenamefont {Lackey}, \citenamefont {Spentzouris},
  \citenamefont {Jungman},\ and\ \citenamefont
  {Spentzouris}}]{amundson2003calibration}%
  \BibitemOpen
  \bibfield  {author} {\bibinfo {author} {\bibfnamefont {J.}~\bibnamefont
  {Amundson}}, \bibinfo {author} {\bibfnamefont {J.}~\bibnamefont {Lackey}},
  \bibinfo {author} {\bibfnamefont {P.}~\bibnamefont {Spentzouris}}, \bibinfo
  {author} {\bibfnamefont {G.}~\bibnamefont {Jungman}},\ and\ \bibinfo {author}
  {\bibfnamefont {L.}~\bibnamefont {Spentzouris}},\ }\bibfield  {title}
  {\bibinfo {title} {Calibration of the fermilab booster ionization profile
  monitor},\ }\href@noop {} {\bibfield  {journal} {\bibinfo  {journal}
  {Physical Review Special Topics-Accelerators and Beams}\ }\textbf {\bibinfo
  {volume} {6}},\ \bibinfo {pages} {102801} (\bibinfo {year}
  {2003})}\BibitemShut {NoStop}%
\bibitem [{\citenamefont {Franchetti}\ \emph {et~al.}(2003)\citenamefont
  {Franchetti}, \citenamefont {Hofmann}, \citenamefont {Giovannozzi},
  \citenamefont {Martini},\ and\ \citenamefont {Metral}}]{franchetti2003space}%
  \BibitemOpen
  \bibfield  {author} {\bibinfo {author} {\bibfnamefont {G.}~\bibnamefont
  {Franchetti}}, \bibinfo {author} {\bibfnamefont {I.}~\bibnamefont {Hofmann}},
  \bibinfo {author} {\bibfnamefont {M.}~\bibnamefont {Giovannozzi}}, \bibinfo
  {author} {\bibfnamefont {M.}~\bibnamefont {Martini}},\ and\ \bibinfo {author}
  {\bibfnamefont {E.}~\bibnamefont {Metral}},\ }\bibfield  {title} {\bibinfo
  {title} {Space charge and octupole driven resonance trapping observed at the
  cern proton synchrotron},\ }\href@noop {} {\bibfield  {journal} {\bibinfo
  {journal} {Physical Review Special Topics-Accelerators and Beams}\ }\textbf
  {\bibinfo {volume} {6}},\ \bibinfo {pages} {124201} (\bibinfo {year}
  {2003})}\BibitemShut {NoStop}%
\bibitem [{\citenamefont {Franchetti}\ \emph {et~al.}(2010)\citenamefont
  {Franchetti}, \citenamefont {Chorniy}, \citenamefont {Hofmann}, \citenamefont
  {Bayer}, \citenamefont {Becker}, \citenamefont {Forck}, \citenamefont
  {Giacomini}, \citenamefont {Kirk}, \citenamefont {Mohite}, \citenamefont
  {Omet} \emph {et~al.}}]{franchetti2010experiment}%
  \BibitemOpen
  \bibfield  {author} {\bibinfo {author} {\bibfnamefont {G.}~\bibnamefont
  {Franchetti}}, \bibinfo {author} {\bibfnamefont {O.}~\bibnamefont {Chorniy}},
  \bibinfo {author} {\bibfnamefont {I.}~\bibnamefont {Hofmann}}, \bibinfo
  {author} {\bibfnamefont {W.}~\bibnamefont {Bayer}}, \bibinfo {author}
  {\bibfnamefont {F.}~\bibnamefont {Becker}}, \bibinfo {author} {\bibfnamefont
  {P.}~\bibnamefont {Forck}}, \bibinfo {author} {\bibfnamefont
  {T.}~\bibnamefont {Giacomini}}, \bibinfo {author} {\bibfnamefont
  {M.}~\bibnamefont {Kirk}}, \bibinfo {author} {\bibfnamefont {T.}~\bibnamefont
  {Mohite}}, \bibinfo {author} {\bibfnamefont {C.}~\bibnamefont {Omet}}, \emph
  {et~al.},\ }\bibfield  {title} {\bibinfo {title} {Experiment on space charge
  driven nonlinear resonance crossing in an ion synchrotron},\ }\href@noop {}
  {\bibfield  {journal} {\bibinfo  {journal} {Physical Review Special
  Topics-Accelerators and Beams}\ }\textbf {\bibinfo {volume} {13}},\ \bibinfo
  {pages} {114203} (\bibinfo {year} {2010})}\BibitemShut {NoStop}%
\bibitem [{\citenamefont {Asvesta}\ \emph {et~al.}(2020)\citenamefont
  {Asvesta}, \citenamefont {Bartosik}, \citenamefont {Gilardoni}, \citenamefont
  {Huschauer}, \citenamefont {Machida}, \citenamefont {Papaphilippou},\ and\
  \citenamefont {Wasef}}]{asvesta2020identification}%
  \BibitemOpen
  \bibfield  {author} {\bibinfo {author} {\bibfnamefont {F.}~\bibnamefont
  {Asvesta}}, \bibinfo {author} {\bibfnamefont {H.}~\bibnamefont {Bartosik}},
  \bibinfo {author} {\bibfnamefont {S.}~\bibnamefont {Gilardoni}}, \bibinfo
  {author} {\bibfnamefont {A.}~\bibnamefont {Huschauer}}, \bibinfo {author}
  {\bibfnamefont {S.}~\bibnamefont {Machida}}, \bibinfo {author} {\bibfnamefont
  {Y.}~\bibnamefont {Papaphilippou}},\ and\ \bibinfo {author} {\bibfnamefont
  {R.}~\bibnamefont {Wasef}},\ }\bibfield  {title} {\bibinfo {title}
  {Identification and characterization of high order incoherent space charge
  driven structure resonances in the cern proton synchrotron},\ }\href@noop {}
  {\bibfield  {journal} {\bibinfo  {journal} {arXiv preprint arXiv:2005.06575}\
  } (\bibinfo {year} {2020})}\BibitemShut {NoStop}%
\bibitem [{\citenamefont {Molodozhentsev}\ \emph {et~al.}(2007)\citenamefont
  {Molodozhentsev}, \citenamefont {Tomizawa},\ and\ \citenamefont
  {Koseki}}]{molodozhentsev2007space}%
  \BibitemOpen
  \bibfield  {author} {\bibinfo {author} {\bibfnamefont {A.}~\bibnamefont
  {Molodozhentsev}}, \bibinfo {author} {\bibfnamefont {M.}~\bibnamefont
  {Tomizawa}},\ and\ \bibinfo {author} {\bibfnamefont {T.}~\bibnamefont
  {Koseki}},\ }\bibfield  {title} {\bibinfo {title} {Space charge effects for
  jparc main ring},\ }in\ \href@noop {} {\emph {\bibinfo {booktitle} {2007 IEEE
  Particle Accelerator Conference (PAC)}}}\ (\bibinfo {organization} {IEEE},\
  \bibinfo {year} {2007})\ pp.\ \bibinfo {pages} {3315--3317}\BibitemShut
  {NoStop}%
\bibitem [{\citenamefont {Ohmi}\ \emph {et~al.}(2014)\citenamefont {Ohmi},
  \citenamefont {Harada}, \citenamefont {Igarashi},\ and\ \citenamefont
  {Sato}}]{ohmi2014study}%
  \BibitemOpen
  \bibfield  {author} {\bibinfo {author} {\bibfnamefont {K.}~\bibnamefont
  {Ohmi}}, \bibinfo {author} {\bibfnamefont {H.}~\bibnamefont {Harada}},
  \bibinfo {author} {\bibfnamefont {S.}~\bibnamefont {Igarashi}},\ and\
  \bibinfo {author} {\bibfnamefont {Y.}~\bibnamefont {Sato}},\ }\bibfield
  {title} {\bibinfo {title} {Study for space charge effect in tune space at
  j-parc mr},\ }in\ \href@noop {} {\emph {\bibinfo {booktitle} {5th Int.
  Particle Accelerator Conf.(IPAC'14), Dresden, Germany, June 15-20, 2014}}}\
  (\bibinfo {organization} {JACOW, Geneva, Switzerland},\ \bibinfo {year}
  {2014})\ pp.\ \bibinfo {pages} {2100--2102}\BibitemShut {NoStop}%
\bibitem [{\citenamefont {Hotchi}\ \emph {et~al.}(2017)\citenamefont {Hotchi},
  \citenamefont {Harada}, \citenamefont {Hayashi}, \citenamefont {Kato},
  \citenamefont {Kinsho}, \citenamefont {Okabe}, \citenamefont {Saha},
  \citenamefont {Shobuda}, \citenamefont {Tamura}, \citenamefont {Tani} \emph
  {et~al.}}]{hotchi2017achievement}%
  \BibitemOpen
  \bibfield  {author} {\bibinfo {author} {\bibfnamefont {H.}~\bibnamefont
  {Hotchi}}, \bibinfo {author} {\bibfnamefont {H.}~\bibnamefont {Harada}},
  \bibinfo {author} {\bibfnamefont {N.}~\bibnamefont {Hayashi}}, \bibinfo
  {author} {\bibfnamefont {S.}~\bibnamefont {Kato}}, \bibinfo {author}
  {\bibfnamefont {M.}~\bibnamefont {Kinsho}}, \bibinfo {author} {\bibfnamefont
  {K.}~\bibnamefont {Okabe}}, \bibinfo {author} {\bibfnamefont
  {P.}~\bibnamefont {Saha}}, \bibinfo {author} {\bibfnamefont {Y.}~\bibnamefont
  {Shobuda}}, \bibinfo {author} {\bibfnamefont {F.}~\bibnamefont {Tamura}},
  \bibinfo {author} {\bibfnamefont {N.}~\bibnamefont {Tani}}, \emph {et~al.},\
  }\bibfield  {title} {\bibinfo {title} {Achievement of a low-loss 1-mw beam
  operation in the 3-gev rapid cycling synchrotron of the japan proton
  accelerator research complex},\ }\href@noop {} {\bibfield  {journal}
  {\bibinfo  {journal} {Physical Review Accelerators and Beams}\ }\textbf
  {\bibinfo {volume} {20}},\ \bibinfo {pages} {060402} (\bibinfo {year}
  {2017})}\BibitemShut {NoStop}%
\bibitem [{\citenamefont {Bhat}()}]{bhat2020esme}%
  \BibitemOpen
  \bibfield  {author} {\bibinfo {author} {\bibfnamefont {C.}~\bibnamefont
  {Bhat}},\ }\href@noop {} {\bibinfo {title} {{$\gamma_t$-jump in Booster
  during PIP2 era}}},\ \bibinfo {howpublished} {Beams-doc-8735-v1},\ \bibinfo
  {note} {(Fermilab internal note)}\BibitemShut {NoStop}%
\bibitem [{\citenamefont {Derwent}()}]{derwent2020blond}%
  \BibitemOpen
  \bibfield  {author} {\bibinfo {author} {\bibfnamefont {P.}~\bibnamefont
  {Derwent}},\ }\href@noop {} {\bibinfo {title} {{Implementation of BLonD for
  Booster Simulations}}},\ \bibinfo {howpublished} {Beams-doc-8690-v1},\
  \bibinfo {note} {(Fermilab internal note)}\BibitemShut {NoStop}%
\bibitem [{\citenamefont {Kapin}\ \emph {et~al.}()\citenamefont {Kapin} \emph
  {et~al.}}]{kapin2017collimation}%
  \BibitemOpen
  \bibfield  {author} {\bibinfo {author} {\bibfnamefont {V.}~\bibnamefont
  {Kapin}} \emph {et~al.},\ }\href@noop {} {\bibinfo {title} {{Study of
  Two-Stage Collimation System in Fermilab Booster}}},\ \bibinfo {howpublished}
  {Beams-doc-5519-v1},\ \bibinfo {note} {(Fermilab internal note)}\BibitemShut
  {NoStop}%
\bibitem [{\citenamefont {Shiltsev}(2015)}]{shiltsev2015electron}%
  \BibitemOpen
  \bibfield  {author} {\bibinfo {author} {\bibfnamefont {V.~D.}\ \bibnamefont
  {Shiltsev}},\ }\href@noop {} {\emph {\bibinfo {title} {Electron lenses for
  super-colliders}}}\ (\bibinfo  {publisher} {Springer},\ \bibinfo {year}
  {2015})\BibitemShut {NoStop}%
\bibitem [{\citenamefont {Danilov}\ and\ \citenamefont
  {Nagaitsev}(2010)}]{danilov2010nonlinear}%
  \BibitemOpen
  \bibfield  {author} {\bibinfo {author} {\bibfnamefont {V.}~\bibnamefont
  {Danilov}}\ and\ \bibinfo {author} {\bibfnamefont {S.}~\bibnamefont
  {Nagaitsev}},\ }\bibfield  {title} {\bibinfo {title} {Nonlinear accelerator
  lattices with one and two analytic invariants},\ }\href@noop {} {\bibfield
  {journal} {\bibinfo  {journal} {Physical Review Special Topics-Accelerators
  and Beams}\ }\textbf {\bibinfo {volume} {13}},\ \bibinfo {pages} {084002}
  (\bibinfo {year} {2010})}\BibitemShut {NoStop}%
\bibitem [{\citenamefont {Antipov}\ \emph {et~al.}(2017)\citenamefont
  {Antipov}, \citenamefont {Broemmelsiek}, \citenamefont {Bruhwiler},
  \citenamefont {Edstrom}, \citenamefont {Harms}, \citenamefont {Lebedev},
  \citenamefont {Leibfritz}, \citenamefont {Nagaitsev}, \citenamefont {Park},
  \citenamefont {Piekarz} \emph {et~al.}}]{antipov2017iota}%
  \BibitemOpen
  \bibfield  {author} {\bibinfo {author} {\bibfnamefont {S.}~\bibnamefont
  {Antipov}}, \bibinfo {author} {\bibfnamefont {D.}~\bibnamefont
  {Broemmelsiek}}, \bibinfo {author} {\bibfnamefont {D.}~\bibnamefont
  {Bruhwiler}}, \bibinfo {author} {\bibfnamefont {D.}~\bibnamefont {Edstrom}},
  \bibinfo {author} {\bibfnamefont {E.}~\bibnamefont {Harms}}, \bibinfo
  {author} {\bibfnamefont {V.}~\bibnamefont {Lebedev}}, \bibinfo {author}
  {\bibfnamefont {J.}~\bibnamefont {Leibfritz}}, \bibinfo {author}
  {\bibfnamefont {S.}~\bibnamefont {Nagaitsev}}, \bibinfo {author}
  {\bibfnamefont {C.-S.}\ \bibnamefont {Park}}, \bibinfo {author}
  {\bibfnamefont {H.}~\bibnamefont {Piekarz}}, \emph {et~al.},\ }\bibfield
  {title} {\bibinfo {title} {Iota (integrable optics test accelerator):
  facility and experimental beam physics program},\ }\href@noop {} {\bibfield
  {journal} {\bibinfo  {journal} {Journal of Instrumentation}\ }\textbf
  {\bibinfo {volume} {12}}\bibinfo  {number} { (03)},\ \bibinfo {pages}
  {T03002}}\BibitemShut {NoStop}%
\end{thebibliography}%
\end{document}